\documentclass[10pt,preprint]{aastex}

\shorttitle{HI and NANTEN GMC in the LMC}
\shortauthors{Fukui et al.}
\begin{document}
\title{Molecular and Atomic Gas in the Large Magellanic Cloud II.
Three-dimensional Correlation between CO and HI}
\author{Fukui, Y.\altaffilmark{1},
Kawamura, A.\altaffilmark{1},
Wong, T.\altaffilmark{2}
Murai, M.\altaffilmark{1}
Iritani, H.\altaffilmark{1}
Mizuno, N.\altaffilmark{1,3},
Mizuno, Y.\altaffilmark{1},\\
Onishi, T.\altaffilmark{1,4},
Hughes, A.\altaffilmark{5,6}
Ott, J.\altaffilmark{7,8}
Muller, E.\altaffilmark{1}
Staveley-Smith, L.\altaffilmark{9} \&
Kim, S.\altaffilmark{10}
}

\altaffiltext{1}{Department of Astrophysics, Nagoya University, Furocho, Chikusaku, Nagoya 464-8602, Japan}
\altaffiltext{2}{Astronomy Department, University of Illinois, 1002 W. Green St, Urbana, IL 61801}
\altaffiltext{3}{National Astronomical Observatory of Japan, 2-21-1 Osawa, Mitaka, Tokyo 181-8588, Japan}
\altaffiltext{4}{Department of Physical Science, Osaka Prefecture University, Gakuen 1-1, Sakai, Osaka 599-8531, Japan}
\altaffiltext{5}{Centre for Astrophysics and Supercomputing, Swinburne University of Technology, PO Box 218, Hawthorn, VIC 3122, Australia}
\altaffiltext{6}{CSIRO Australia Telescope National Facility, PO Box 76, Epping, NSW 1710, Australia}
\altaffiltext{7}{National Radio Astronomy Observatory, 520 Edgemont Rd, Charlottesville, VA 22903}
\altaffiltext{8}{California Institute of Technology, MC 10524, Pasadena, CA 91125}
\altaffiltext{9}{School of Physics M013, University of Western Australia, Crawley, WA 6009, Australia}
\altaffiltext{10}{Department of Astronomy and Space Science, Sejong University, KwangJin-gu, KunJa-dong 98, Seoul 143-747, Korea}
\email{fukui@a.phys.nagoya-u.ac.jp, kawamura@a.phys.nagoya-u.ac.jp, murai@a.phys.nagoya-u.ac.jp}

\begin{abstract}
We compare the CO ($J =$1--0) and HI emission in the Large Magellanic
Cloud (LMC) in three dimensions, i.e. including a velocity axis in
addition to the two spatial axes, with the aim of elucidating the
physical connection between giant molecular clouds (GMCs) and their
surrounding HI gas. The CO $J =$1--0 dataset is from the second NANTEN
CO survey and the HI dataset is from the merged Australia Telescope
Compact Array (ATCA) and Parkes Telescope surveys. The major findings
of our analysis are: 1) GMCs are associated with an envelope of HI
emission, 2) in GMCs [average CO intensity] $\propto$ [average HI
  intensity]$^{1.1\pm0.1} $ and 3) the HI intensity tends to increase
with the star formation activity within GMCs, from Type I to Type
III. An analysis of the HI envelopes associated with GMCs shows that
their average linewidth is 14 km s$^{-1}$ and the mean density in the
envelope is 10 cm$^{-3}$. We argue that the HI envelopes are gravitationally bound by GMCs. These findings are consistent
with a continual increase in the mass of GMCs via HI accretion at an
accretion rate of 0.05 $M_{\sun}$yr$^{-1}$ over a time scale of 10
Myr. The growth of GMCs is terminated via dissipative ionization
and/or stellar-wind disruption in the final stage of GMC evolution.
\end{abstract}
\keywords{Magellanic Clouds --- ISM: clouds --- ISM: molecules --- 
ISM: atoms ---galaxies: ISM --- radio lines: ISM }

\section{Introduction}

Giant molecular clouds (GMCs), the most massive aggregations of
interstellar matter with $10^{5-6} M_{\sun}$, are the principal sites
of star formation in galaxies. It is important to understand how GMCs
are formed out of the less dense atomic interstellar gas in order to
understand galactic evolution. The interstellar HI gas has densities
of less than several 10 cm$^{-3}$ while molecular clouds have
densities larger than 100 cm$^{-3}$. It is reasonable to assume that
HI is being converted into H$_{2}$ either by thermal/gravitational
instabilities and/or shock compressions, although the detailed
processes of this conversion are not yet well-understood. Sato \&
Fukui (1978) and Hasegawa, Sato, \& Fukui (1983) identified cold HI
gas associated with GMCs in M17 and W3/4 and suggested that the cold
HI gas may be converted into molecular gas for these
GMCs. Subsequently, Wannier et al (1983) showed that five molecular
clouds are associated with warm HI envelopes and suggested that such
HI envelopes may be common around GMCs. Nonetheless, associations
between GMCs and HI envelopes are difficult to identify systematically
throughout the Galactic disk, since GMCs samples are restricted to the
solar vicinity due to the crowding effects (Andersson, Wannier, \&
Morris 1991). As a consequence, the GMC-HI association has not been well
established.

The Magellanic system -- including the LMC, the SMC and the Bridge --
is an ideal laboratory to study star formation and molecular cloud
evolution because of its proximity to the Milky Way (e.g., Fukui et al
1999; Mizuno et al. 2001; Mizuno et al. 2006; Fukui et al. 2008; Ott
et al. 2008; Kawamura et al, 2009). We expect that the Magellanic
system can also shed light on the physical connection between GMCs and
their atomic surroundings. Indeed, the LMC may offer the best place
for such a study because of its nearly face-on orientation and level
of star formation activity.  The LMC's molecular cloud population,
which is best traced via the CO emission, provides a key to understand
the galaxy's star formation. Molecular clouds are able to highlight
the location of star formation due to their highly clumped
distribution in both space and velocity. The LMC's atomic gas, by
contrast, has lower densities and is only weakly coupled to sites of
active star formation, but it is the most promising candidate for the
mass reservoir of GMC formation (e.g., Blitz et al. 2007). We note,
moreover, that cold HI gas has been detected in the LMC (e.g., Dickey
et al. 1994), and that the correlation between CO and HI may provide
crucial observational evidence about the molecular cloud formation
process.

Wong et al. (2009) compared the HI and CO emission throughout the LMC
on a pixel-by-pixel basis using the second NANTEN CO and ATCA+Parkes
HI datasets. These authors studied correlations between the integrated
CO and HI intensities, where the latter was integrated over all
velocities with HI emission or over individual Gaussian
components. They found that CO emission is associated with high
intensity HI gas but that intense HI emission is not always associated
with CO. They also discovered a weak tendency for CO to be associated
with HI components that have relatively low velocity dispersion. This
suggests that energy dissipation of the HI gas may be required for the
formation of molecular clouds.  Following the global analysis by Wong
et al. (2009), we focus here on the HI associated with individual GMCs
in the LMC. In order to address this issue, we conduct a detailed
comparison between the CO and HI emission in three dimensions,
i.e. $(x,y,v)$, at a spatial resolution of $\sim$40 pc and a velocity
resolution of 1.7 km ss$^{-1}$. The present study is complementary to
the work by Wong et al. (2009); in conjunction, the two studies
provide a new insight into the CO-HI connection. In section 2, we
briefly review the basic observational properties of GMCs in the
LMC. In section 3, we describe our method of analysing the CO-HI
correlation and present our results. We discuss the physical
interpretation of our results in section 4 and provide a summary of
our major conclusions in section 5.
 
\section{GMCs in the LMC; the second NANTEN CO survey}

The LMC is extended by more than 30 square degrees across the sky and
it has been a difficult task to make a systematic survey of the CO
emission at angular resolutions sufficient to resolve individual
GMCs. Fukui et al. (1999) made such a survey in the 2.6mm CO line with
the NANTEN 4m mm-wave telescope and published the first results in
Fukui et al. (1999). Subsequently, these authors completed another
survey of the LMC, improving the sensitivity by a factor of two. The
second NANTEN CO survey has cataloged 272 GMCs (Fukui et
al. 2008). The basic physical parameters of GMCs in the LMC are
similar to those in the Milky Way and other nearby galaxies. Their
masses range from $\sim 10^{5} M_{\sun}$ to $\sim 10^{6} M_{\sun}$;
the mass spectrum is quite steep with a slope of $dN/dM \sim
M^{-2}$. The X factor, the ratio of the H$_{2}$ column density to CO
intensity, is $\sim 7 \times 10^{20}$ cm$^{-2}$ (K km s$^{-1}$)
$^{-1}$ (Fukui et al. 2008; Blitz et al. 2007). The complete sampling
of the NANTEN survey has also allowed us to make a statistical study
of GMCs with various young objects including HII regions and young
stellar clusters. Kawamura et al. (2009) confirmed that there are
three classes of GMCs, that can be categorised according to their
association with young star clusters as originally indicated in the
analysis of the first NANTEN LMC survey (Fukui et al. 1999; Yamaguchi
et al. 2001). Type I GMCs show no signs of active star formation in
the sense that no O stars are being formed. Type II GMCs are
associated with small HII region(s), indicating the formation of
isolated O stars, but do not host any stellar clusters identified by
Bica et al. (1996). Type III GMCs are actively forming stars as shown
by their association with large HII regions and young stellar
clusters. These classes are interpreted as an evolutionary sequence
from Type I to III; the lifetime of a GMC is estimated to be a few
$\times$ 10 Myr in total (Kawamura et al. 2009). The stage after Type
III is likely to be a very violent dissipation of the GMC due to the
UV photons and stellar winds produced by the nascent clusters, as seen
spectacularly in the region of 30 Dor (Yamaguchi et al. 2001). A
comparison of the physical parameters of the GMCs shows that the size
and mass of the clouds tend to increase from Type I/II to Type III. A
summary of the average mass and size for the three GMC classes is
presented in Table 1, as taken from Kawamura et al. (2009).
 
\section{Correlation between CO and HI}
\subsection{3-D correlation}

Previous studies of star formation in galaxies have employed
2-dimensional (2-D) maps of HI intensity with large spatial averaging
on scales between $\sim$100 pc and 1 kpc (e.g., Schmidt 1972; Kennicut et
al. 1988). Here, we make a 3-D comparison between the CO and HI in the
LMC where the 3-D datacubes have a velocity axis in addition to two
spatial axes projected on the sky. Preliminary results of the
comparison have been published elsewhere (Fukui 2007). We use the 3-D
datacube of CO obtained with NANTEN (Fukui et al. 2008) and an HI
datacube obtained with ATCA and Parkes (Kim et al. 2003). The CO
emission traces GMCs and the HI emission traces less dense atomic
gas. Figure 1 shows an overlay of the velocity-integrated intensities
of the CO and HI emission; from this, it is clear that GMCs in the LMC
tend to be located towards HI filaments or local HI peaks, suggesting
that HI is a prerequisite for GMC formation (Blitz et al. 2007).
However, it is also clear that there are many HI peaks and filaments
without CO emission (Wong et al. 2009). Figure 2 shows typical CO and
HI line profiles in the LMC. The CO emission is highly localized in
velocity: the HI emission ranges over 100 km s$^{-1}$ while the CO
emission has a typical linewidth of less than 10 km s$^{-1}$. We note
that the large velocity dispersion of HI may be dominated by
physically unrelated velocity components along the line of sight,
i.e. the HI gas associated with the GMC may only be the small fraction
of HI with velocities close to that of the CO emission. Previous
studies of the CO-HI connection that use velocity integrated 2-D maps
may therefore overestimate the intensity of the associated HI emission
along each line of sight. By making use of the velocity dimension, the
present 3-D analysis may allow us to identify the HI gas that is
physically connected to the GMCs. The NANTEN and ATCA+Parkes datacubes
have somewhat different spatial and velocity resolutions, so we have
convolved both datasets to a spatial resolution of 40 pc $\times$ 40
pc, and a velocity resolution of 1.7 km s$^{-1}$. The total number of
3-D pixels is approximately $2 \times 10^{6}$ across the area of
surveyed by NANTEN. The HI and CO intensities are expressed in units
of $T_{\rm b}$(K) and $T^{*}_{\rm R}$ (K); the $3 \sigma$ noise levels
of the HI and CO datacubes are 7.2 K and 0.21 K respectively. Strictly
speaking, the one-to-one correspondence between a velocity and a
position is not guaranteed because there is a chance that physically
unrelated HI gas may have the same velocity as HI gas related to the
GMC along the same line of sight. Our results identify the HI
associated with GMCs and suggest that such contamination along
individual sightlines may not be a serious problem. We further note
that HI absorption towards background radio continuum sources does not
affect significantly the HI intensity at the present spatial
resolution, as verified toward 30 Dor, one of the brightest radio
continuum sources in the LMC.

Figure 3 shows a histogram of the HI intensity in the 3-D
datacube. Pixels with the significant CO emission ($T^{*}_{\rm R} > $
0.21 K) are shown in red. Throughout this paper, histograms always use
the values of 3-D pixels with no spatial integration unless otherwise
stated. The histogram in Figure 3 shows that the fraction of
CO-detected pixels increases monotonically with the HI intensity,
suggesting the HI intensity is a necessary condition to form GMCs,
consistent with the conclusion by Wong et al. (2009). About one third
of the pixels with $T_{\rm b}$(HI) of $\sim$ 90 K exhibit CO emission,
but it seems that there is no sharp threshold value of HI intensity
that is required for GMC formation.

Figure 4 shows a histogram of the HI intensity for the three GMC
types. Each pixel detected in CO belongs to one of the GMCs cataloged
in Fukui et al (2008). Figure 4 clearly shows that the HI intensity
tends to increase from Type I to Type III, although the dispersion is
considerable. The average HI intensity for Types I, II and III GMCs
across the LMC is $34 \pm 16$ K($1 \sigma$), $47 \pm 17$ K and $56 \pm
19$ K, i.e. the average HI intensity increases with the level of star
formation activity within the GMC. In order to test for variation
within the galaxy, we tentatively divide the galaxy into three
regions, i.e., Bar, North and Arc, as shown in Figure
4(right). Histograms for each region, shown in the lower three panels
of Figure 4, reveal the same trend, suggesting that the present trend
is common over the whole LMC. The total number of the pixels in each
region is 429,510, 666,060, and 405,093. Type I, Type II and Type III
GMCs include 330, 1,065, and 639 pixels in the Bar region; 346, 878,
and 1,158 pixels in the North region; and 389, 650, and 1,231 pixels
in the Arc region.

Figure 5 shows the velocity channel maps of the HI distribution
associated with Type I, Type II, and Type III GMCs. These channel maps
clearly show that the HI is associated with the CO. The CO
distribution has small structures of $\sim$ 100 pc or less and the HI
appears to be associated with the GMC on larger scales of $\sim$ 100
to 400 pc. The HI emission is not always symmetric with respect to a
GMC, even though HI typically envelopes each GMC. The associated HI is
often elongated along the GMCs and the region of intense HI emission
is usually $<$ 100 pc wide. The CO emission typically extends over a
velocity range of $\sim 5$ km s$^{-1}$; beyond a few times this
velocity range, the associated HI emission generally becomes much
weaker or disappears.
 
\subsection{Physical properties of the HI envelope}

In general, it is a complicated task to derive reliable physical
properties of the HI gas associated with a GMC because the HI profiles
are a blend of several different components along the line of sight,
making it difficult to select the HI gas that is physically connected
to a GMC. Another obstacle is that the HI emission is spatially more
extended than the CO emission and has a less clear boundary than the
CO.
 
For our analysis, we first selected GMCs with simple single-peaked HI
profiles from the Fukui et al (2008) catalog. The resulting sample
consists of 123 GMCs in total. Their catalog numbers and basic
physical properties, taken from Fukui et al. (2008), are listed in
Table 2. For these GMCs, we tested whether there was a bias in their
location with respect to the kinematic center of the galaxy, in their
CO linewidth or in their molecular mass. The histograms in Figure 6
indicate that there is no particular trend for these properties of the
selected GMCs compared to GMCs in the complete catalog, suggesting
that there is no appreciable selection bias. We applied a
Kolmogorov-Smirnov test to the three histograms and calculated maximum
deviations of 0.031, 0.061 and 0.117 respectively for the three
parameters. These values are less than the critical deviation, 0.129,
for a conventional significance level of 0.05, confirming that there
is no selection bias.

Next, we made Gaussian fits to the HI and CO profiles towards the CO
peak of each GMC. This procedure yields a peak intensity, peak
velocity and half-power linewidth for each line profile (a summary is
given for each GMC type in Table 1). Figure 7 shows the relation
between the CO linewidth and the difference between the CO and HI peak
velocities. We find the HI and CO peak velocities to be in good
agreement, showing only a small scatter of less than a few km
s$^{-1}$. Figure 8 shows two histograms of the HI and CO
linewidths. We see that the HI linewidth is typically 14 km s$^{-1}$,
roughly three times larger than that of CO.  Figure 9 shows a
correlation between HI and CO linewidths. The two quantities show a
positive correlation with a correlation coefficient of 0.39. The
correlation coefficient is determined using the Spearman rank method
throughout the present paper. The kinematic properties of HI and CO,
as illustrated in Figures 7 and 9, lend further support to a physical
association between the HI and CO.

In order to estimate the size of the HI envelope surrounding each GMC,
we construct an HI integrated intensity map of each GMC. First, we
find the local peak in the HI intensity cube surrounding the CO
emission, and then integrate the HI intensity over the velocity
channels corresponding to the FWHM of the HI line profile at this peak
position. Next we estimate the area, $S$, where the HI integrated
intensity is greater than 80 \% of the value at the local HI peak. We
then calculate the radius of the HI envelope, $R$(HI) from its
projected area, $S = \pi R({\rm HI})^{2}$. The HI integrated intensity
is calculated for all the pixels with detectable CO emission; the
spatial distribution of the HI emission generally shows a peak and a
reasonably defined boundary. The 80 \% level was chosen after a few
trials using different levels; it is the maximum value for which a
reasonable HI size is obtained for 116 of the 123 envelopes. While 80
\% seems to be rather high for such a definition of a cloud envelope,
the HI size can be unrealistically large compared to the CO cloud size
along a filamentary HI distribution if we use a lower level such as 60
\%; an example is shown in Figure 5b. The HI radius is then corrected
for beam dilution by adopting Gaussian de-convolution with an HI FWHM
of 2.6 arcmin, i.e. the same procedure that is applied to the radius
of the GMC determined from the CO emission, $R$(CO) (Kawamura et al
2009). $R$(CO), the CO peak position, $R$(HI), the HI peak position,
and the deviation between the peaks in parsecs, $\sqrt{(\Delta
  \alpha^{2}+\Delta \delta^{2})}$, are listed in Table 2. Seven
clouds, for which $R$(HI) comprises only a few pixels, are denoted by
asterisks. The physical extent of the HI envelopes can be as large as
a few 100 pc. Ws long as we use the local HI peak of individual GMCs
as our reference point, it seems difficult to contrive an alternative
uniform definition of the envelope size is difficult using the present
HI dataset.

In Figure 10, we show a histogram of the spatial deviation of the HI
and CO peaks. This shows that nearly 80 \% of the HI envelopes peak
within 120 pc of the local CO peak, and that nearly 60 \% of the HI
envelopes peak within 80 pc of the CO peak.  These separations may
seem large compared to $R$(HI), but we argue that they are reasonable
if the HI is enveloping CO at scales over $\sim$ 300 pc. It should be
noted that the HI envelopes are not concentric with the CO emission,
but are rather ``enveloping'' with some offsets in peak positions as
illustrated in Figure 5. We thus expect to find some difference in
general between the peak positions of the CO and HI emission (as seen
in Figure 10), but the fact that the majority of HI peaks are located
within 120 pc of the CO peaks is clearly suggestive of a physical
association between the GMCs and their surrounding atomic gas. In
Figure 11, we show a correlation between $R$(HI) and $R$(CO) for 62
GMCs whose radius is greater than 30 pc and find that they are
positively correlated with a correlation coefficient of 0.45. Despite
the relatively flatter distribution of the HI, the size of the HI
envelope does seem to correlate with the size of the GMC.

To summarize our analysis in this section, we find for the 123 GMCs
with single-peaked HI profiles that: 1) the peak velocities of the CO
and HI are in good agreement (Figure 7); 2) the CO and HI linewidths
show a positive correlation (Figure 9); 3) the HI envelopes, defined
using the 80 \% level of the local HI integrated intensity peak, are
mostly ($\sim 80$ \%) centred within 120 pc of the peak CO position
(Figure 10); and 4) the radius of the HI envelope is positively
correlated with the size of the GMC for GMCs with radii greater than
30 pc (Figure 11). These four results lend further support to the idea
that HI envelopes are physically associated with GMCs, reinforcing the
impression conveyed by a global comparison between the HI CO emission
in the LMC (Figure 3) and the morphological similarity between the CO
and HI in individual velocity channels. (Figure 5).

Next, we made an estimate of the HI column density for the 123 GMCs by
using the relation ($N$ (HI) [cm$^{-2}$] $= 1.8 \times 10^{18}\int
T_{\rm b} dv$[K km s$^{-1}$]). The average values for the three GMC
types are listed in Table 1. We find the peak HI column density is
mostly in a range of 2--5 $\times 10^{21}$ cm$^{-2}$. We estimate the
typical density in the HI envelopes to be $\sim 10$ cm$^{-1}$ by
dividing the peak HI column density 2--5 $\times 10^{21}$ cm$^{-2}$ by
the typical size of the associated HI 50--100 pc (see Figure 11). The
mass of the HI envelopes is large, typically $\sim 5 \times 10^{4}
M_{\sun}$. The HI envelopes are likely gravitationally bound by GMCs because a half of the HI line width, 7 km s$^{-1}$, is nearly equal to $\sqrt{GM/R} \sim 6$ km s$^{-1}$ for $M = 2 \times 10^5 M_{\sun}$ and $R= 40$ pc, the average values of Type II GMCs.

In Figure 12, we plot the relationship between the average CO and HI
luminosity of the 123 GMCs. For each GMC, we selected pixels where CO
emission is significantly detected: only these pixels are used to
calculate the HI and CO luminosity of each cloud. In order to derive
the average CO luminosity of a GMC, we estimated $I$(CO) K km s$^{-1}$ by
dividing the sum of all the significant CO emission within the GMC by
the projected area of the GMC. $I$(HI) K km s$^{-1}$ is calculated in a
similar manner, summing up the HI emission from the pixels within the
GMC with significant CO emission and dividing the result by the area
of the GMC. The regression shown in Figure 12 is well fitted by a power
law with an index of 1.1, indicating a nearly linear correlation
between $I$(CO) and $I$(HI) in a GMC.
 
\section{Discussion}
\subsection{GMCs with HI envelopes}

The present analysis has successfully identified the HI envelopes
associated with GMCs on the basis of a 3-D analysis of GMCs in the
LMC. The HI intensity in the envelope depends on the star-forming
activity within the GMC in the sense that the integrated HI intensity
in the envelope increases from Type I to Type III (Figure 4, section
3.1). In other words, massive GMCs have massive HI envelopes and less
massive GMCs have less massive HI envelopes.

The HI intensity is a product of the spin temperature and the optical
depth of the HI 21 cm transition, provided that the line is optically
thin. This is likely to be the case as the HI profiles show few hints
of saturation like a flat top. The observed maximum HI brightness
temperature is around 100 K and this suggests that the HI spin
temperature is significantly higher than 100 K. Therefore we infer
that the HI intensity should represent optical depth and, accordingly,
HI column density, if the spin temperature is roughly uniform across
the LMC. The HI spin temperature in the LMC may be higher than in the
Galaxy due to a more intense UV field and the lower dust extinction
(Israel 1996; Imara \& Blitz 2007; Dobashi et al. 2008). For the sake
of discussion, we shall tentatively assume that the spin temperature
lies between 150--600 K and is fairly uniform in the HI envelope. We
note that the HI mass is accurately determined as long as the HI
emission is optically thin. The HI mass does not depend on Ts under
the optically thin assumption, because the level populations of the
spin doublet having only $\sim 10^{-5}$ eV is well thermalized in any
realistic density range due to the slow magnetic dipole decay in $\sim
10^7$ yrs.

Fukui et al. (1999) suggest that the three classes of GMC indicate an
evolutionary sequence from Type I to Type III in a few 10 Myr (instead
of ``Type'', these authors used ``Class'' with the same
meaning). Kawamura et al. (2009) present a more detailed analysis of
the association between GMCs and young stellar clusters, confirming
Fukui et al.'s evolutionary scheme. These studies indicate that only
the youngest star clusters with an age less than $\sim 10$ Myr are
clearly associated with GMCs, and that older clusters with an age
greater than 10 Myr are not associated with GMCs. Assuming a steady
state scenario, this implies that the natal gas of clusters is quickly
disrupted within 10 Myr. Considering the complete sampling of both
clusters and GMCs in the LMC, this strongly suggests that the
population of Type III GMCs must be replenished on timescales of 10
Myr. Since the typical time scale of GMC formation is at least 10 Myr,
as estimated by the crossing timescale -- i.e., the cloud size divided
by its velocity dispersion, 100 pc$/$10 km/s $= 10$ Myr, a measure of
the minimum timescale for GMC formation -- we expect to have a similar
population of Type III GMCs and Type III precursors. A straightforward
interpretation is that Type I and Type II GMCs are these precursors
(see for details Kawamura et al. 2009). An alternative possibility is
a more ad-hoc situation in which Type III GMCs are formed suddenly in
a few Myr by an external disturbance, such as a dynamical
interaction. Such a strongly time-dependent scenario seems unlikely,
however, since the three GMC types are fairly uniformly distributed
over the LMC (Kawamura et al. 2009). Figure 4 of this paper also
shows that the three classes of GMCs are distributed across the
galaxy.

We have also seen that Type III GMC tend to be more massive than Type
I and Type II GMCs (Table 2; Kawamura et al. 2009). A natural
interpretation within the evolutionary scenario is that the enveloping
HI gas accretes onto a GMC and that the GMC mass increases with
time. The accreted HI envelope is converted into H$_{2}$ in $\sim 10$
Myr due to increased density, optical extinction and UV
shielding. This infall scenario is consistent with the linear
relationship between $I$(HI) and $I$(CO) in Figure 12; by contrast,
growth of GMCs via collisions between HI clouds would have a steeper
relationship, with $I$(CO) proportional to $I$(HI)$^{2}$. For an
infall scenario, the infall motion can arise from the gravity of a GMC
and possibly from a converging flow driven by super bubbles, while the
thermal motion is negligibly small ~ 1.4-3 km s$^{-1}$ for kinetic
temperatures of $\sim$ 150--600 K. We can roughly estimate the infall
velocity to be half of the HI linewidth, i.e. $\sim 7$ km
s$^{-1}$. This value is consistent with the free fall velocity, $\sim
6$ km s$^{-1}$, for a typical Type II GMC. For spherical accretion, where the GMC is surrounded
by an HI envelope with a radius of $\sim$40 pc, volume density of
$n$(HI) $\sim $10 cm$^{-3}$, and an infall speed of $\sim$ 7 km
s$^{-1}$, we estimate the mass accretion rate to be $\sim$ 0.05
$M_{\sun}$ yr$^{-1}$. Over the typical timescale of the GMC evolution,
i.e. $\sim$ 10 Myr, the increase in molecular mass amounts to $\sim 5
\times 10^{5} M_{\sun}$, which is roughly consistent with the observed
typical value for the mass of a Type III GMC ($\sim 4 \times 10^{5}
M_{\sun}$, Table 1). In the evolutionary picture, the mass accretion
of a GMC is terminated by the violent disruption and/or ionization of
the molecular material by stellar winds and ionization from young
stars.

The infall scenario offers a reasonable interpretation of the HI and
CO properties of GMCs that we have explored in this paper. It remains
to be seen, however, if an infall velocity field is consistent with
2-D observations; careful analysis of an isolated HI envelope with
little kinematical disturbance or nearby contamination could be used
to verify this. It is also important to clarify whether the HI
linewidth is affected by turbulence to a significant degree.

It could be argued that the HI gas surrounding GMCs is supplied by the
recombination of HII into HI, as both Type II and Type III GMCs are
associated with HII regions. This alternative seems unlikely, however:
first, the HII regions in Type II GMCs are compact and therefore do
not constitute a significant mass reservoir; second, the HI envelope
in Type III GMCs are not spatially well-matched with the HII regions
and clusters.  In N159, for instance, the HII regions and young
clusters are confined to the north of the GMC, wherease the HI is more
widely distributed in the east and south (Figure 13).
 
\subsection{HI-H$_{2}$ conversion in GMCs}

This study has shown that the HI and CO distributions correlate well
on 40--100 pc scales. It is worth noting, however, that the
correlation becomes less clear on smaller scales of $\sim 10$ pc
within a GMC. Figure 13 shows an overlay of the HI and CO
distributions for the Type III GMC N159, where the CO data was
obtained using the ASTE 10 m sub-mm telescope in the $^{12}$CO $J
=$3--2 emission line (Mizuno et al., 2008). Figure 13 shows that the
HI becomes less bright at $T_{\rm b}$(HI) 70--80 K toward N159E at
(R.A., Dec.)=($5^{\rm h}40^{\rm m}$, $-69\arcdeg, 45\arcmin$),
compared to intensities of $T_{\rm b}$(HI) $\sim$ 120 K in the HI
envelope. A similar behavior was noted by Ott et al. (2008). This is
unlikely to be cuased by absorption of the radio continuum emission,
as there is no radio continuum emission toward N 159E. We regard this
behaviour to be illustrative of the conversion of HI into H$_{2}$, as
well as the the lower spin temperature in the interior of a GMC. In
the inner part of a GMC, HI is converted into H$_{2}$ via reactions on
grain surfaces on a time scale of $\sim$10 Myr. The HI density is
typically $\sim$1 cm$^{-3}$, compared to a total molecular density of
a few 100 cm$^{-3}$, corresponding to atomic to molecular hydrogen
ratio of $\sim$100 (e.g., Allen \& Robinson 1977; Spitzer 1978;
Goldsmith et al. 2007).  The spin temperature is also lower, and is
likely to equal the molecular gas kinetic temperature of $\sim$ 60 K
(e.g., Sato \& Fukui 1978; Mizuno et al. 2009). In the HI envelope, on
the other hand, the spin temperature is probably between $\sim$
150-600 K and the HI density is estimated to be $\sim$ 10 cm$^{-3}$
with no H$_{2}$. The lower $T_{\rm b}$(HI) in the interior is likely
due to the lower spin temperature and the lower HI density. The mass
of the apparently cold HI gas towards N159E is approximately 10 \% of
the mass of the HI envelope, $\sim 10^{5} M_{\sun}$, if we assume that
the cold HI is optically thin, which suggests that the cold HI within
the GMC is not a dominant mass component of the atomic+molecular cloud
complex.

It has been shown that there are cold HI components in the LMC as
measured from emission and absorption observations toward radio
continuum sources (Dickey et al. 1994). These authors detected HI
absorption features toward 19 of 30 continuum sources in the LMC and
argued that $T_{\rm s}$ of the cold components can be as low as 40
K. Such cold HI components may be associated with GMCs. It is however
not clear observationally how the cold HI in absorption is related to
GMCs because none of the absorption measurements by Dickey et al
(1994) coincide with the NANTEN GMCs.

An issue which has been raised in Wong et al. (2009) is that higher HI
intensity is a necessary but not sufficient condition for CO
formation. In other words, there are many places with high HI
intensities in the LMC without CO. The current analysis has focussed
solely on the HI gas surrounding GMCs and therefore does not directly
address this issue. We note, however, that it is possible that HI gas
with the same intensity can have a significantly different density. We
propose an interpretation that the atomic gas in regions with high HI
intensities but no CO may have lower densities and higher
temperatures. This interpretation could be tested in the future by
high velocity-resolution HI observations that can resolve subtle
variations in HI profiles and hence identify spatial variations of the
atomic gas temperature.

\section{Summary}

We have carried out the first 3-D analysis of the connection between
the CO and HI emission in a galaxy. The major results of our study are
as follows:

\begin{enumerate}
\item A 3-D comparison at a resolution of 40 pc $\times$ 40 pc
  $\times$ 1.7 km s$^{-1}$ has revealed that the fraction of HI
  associated with CO tends to increase as a monotonic function of HI
  intensity without a sharp threshold for the CO formation.

\item We find that GMCs are associated with HI envelopes on scales of
  $\sim$ 50--100 pc. The HI envelopes have typical volume densities of
  $\sim$ 10 cm$^{-3}$ and an average linewidth of $\sim$14 km
  s$^{-1}$, which is about three times larger than the linewidth of
  CO. We argue that the HI envelopes are gravitationally bound by GMCs.

\item For 123 GMCs with single-peaked HI profiles, we find a
  correlation such that [average CO intensity] $\propto$ [average HI
    intensity]$^{1.1 \pm 0.1}$. There is a clear increase of the
  associated HI intensity from GMC Type I to Type III.

\item We interpret our results to mean that a GMC increases in mass
  via continuous HI accretion over a timescale of $\sim$10 Myr and
  with a mass accretion rate of 0.05 $M_{\sun}$ yr$^{-1}$, before
  being disrupted by ionization and stellar winds from young
  clusters. The accreted HI is likely to be converted to molecular
  hydrogen due to the higher shielding within a GMC.
\end{enumerate}

\acknowledgments The NANTEN project is based on a mutual agreement
between Nagoya University and the Carnegie Institution of Washington
(CIW). We greatly appreciate the hospitality of all the staff members
of the Las Campanas Observatory of CIW.  We are thankful to the many
Japanese public donors and companies who contributed to the
realization of the project. This study has made use of SIMBAD
Astronomical Database and NASA's Astrophysics Data System
Bibliographic Services. This work is financially supported in part by
a Grant-in-Aid for Scientific Research from the Ministry of Education,
Culture, Sports, Science and Technology of Japan (No. 15071203), from
JSPS (No. 14102003, No. 18684003, and core-to-core program 17004), and
the Mitsubishi Foundation.

%\clearpage

\clearpage

%\documentclass[preprint]{aastex}
%\begin{document}
%\pagestyle{empty}
\begin{deluxetable}{lcccccc} 
\tablenum{1}
\tablecolumns{7} 
%\tablewidth{0pc} 
\small
\tabletypesize{\scriptsize}
\tablecaption{Physical properties of GMCs} 
\tablehead{
\colhead{GMC Type} &
\colhead{Number of} &
\colhead{$M_{\rm CO}$\tablenotemark{a}} &
\colhead{$R$\tablenotemark{a}}&
\colhead{$N$(HI)\tablenotemark{b}}&
\colhead{$\Delta V_{\rm LSR}$ (HI)\tablenotemark{b}}&
\colhead{$\Delta V_{\rm LSR}$ (CO)\tablenotemark{a}}\\
& 
\colhead{GMCs}&
\colhead{($\times 10^{5} M_{\sun}$)}&
\colhead{(pc)}&
\colhead{($\times 10^{21}$ cm $^{-2}$)}&
\colhead{(km s$^{-1}$)}&
\colhead{(km s$^{-1}$)}}
\startdata 
%\multicolumn{6}{c}{Average properties of the GMCs}\\
%\cline{1-6}\\
Type I  &72 &     \phn2 (2)  &  \phn37 (16)  &  \nodata  &  \nodata  &  \phn5.0 (2.5)  \\
Type II &142 &  \phn2 (3) &  \phn33 (19)  &  \nodata  &  \nodata  &  \phn4.8 (2.2)  \\
Type III &58 &  \phn5 (10)  &  \phn51 (36)  &  \nodata  &  \nodata  &  \phn6.9 (3.0)  \\
Type I (selected)\tablenotemark{c} & 24&   \phn2 (3)  &  \phn35 (17)  &  2.4 (0.9)  &  \phn13.9 (4.0)  &  \phn 4.5 (2.1)  \\
Type II (selected)\tablenotemark{c} &67&  \phn2 (3) &  \phn41 (22)  &  2.6 (1.2) &  \phn14.6 (4.1) &  \phn4.4 (1.6)  \\
Type III (selected)\tablenotemark{c} &32&   \phn4 (3)  &  \phn55 (23)  &  3.3 (1.5)  &  \phn16.1 (3.3)  &  \phn5.5 (1.5) \\
%\cline{1-6}\\
\enddata
\tablenotetext{a}{Fukui et al.\ (2008); Kawamura et al.\ (2009)}
\tablenotetext{b}{Half intensity full width derived by gaussian fitting.}
\tablenotetext{c}{Selected clouds having single peaked HI profiles.}

\tablecomments{Average propeties of the GMCs. The values in parentheses are the standard deviation.}
\end{deluxetable} 

%\end{document}  

\clearpage

\pagestyle{empty}
\begin{deluxetable}{cccccccccccc}

\rotate
\tablenum{2}
\tablecolumns{12} 
\tabletypesize{\footnotesize}
\tablewidth{0pc} 
\tablecaption{List of selected 123 GMCs} 
\tablehead{ 

\colhead{} & \colhead{} & \colhead{} &  \multicolumn{2}{c}{peak position(CO)\tablenotemark{c}} & \colhead{$R$(CO)\tablenotemark{a}} & 
\multicolumn{2}{c}{peak position(HI)\tablenotemark{c}} & \colhead{$R$(HI)\tablenotemark{d}} & \colhead{$N$(HI)\tablenotemark{e}} & \colhead{$\sqrt{\Delta \alpha{^2} + \Delta \delta{^2}}$\tablenotemark{f}} & \colhead{} \\ 
\cline{4-5} \cline{7-8} 

\colhead{number\tablenotemark{a}} & \colhead{name\tablenotemark{a}} & \colhead{Type\tablenotemark{b}} &
\colhead{$\alpha(B1950)$} & \colhead{$\delta(B1950)$}& \colhead{[pc]} & 
\colhead{$\alpha(B1950)$} & \colhead{$\delta(B1950)$}& \colhead{[pc]} & \colhead{$10^{21}$ [cm$^{-2}]$} & \colhead{[pc]} & \colhead{comment\tablenotemark{g}}         }

\startdata 

1 &LMC N J0447-6910	&I	&$\mathrm{	4	h	47.7	m	}$&$\mathrm{	-69	^{\circ }	14 	\arcmin	}$&	44 	&$\mathrm{	4	h	47.7	m	}$&$\mathrm{	-69	^{\circ }	14	\arcmin	}$&	34 	&	2.6 	&	0 	&		\\
4 &LMC N J0449-6910	&III	&$\mathrm{	4	h	49.1	m	}$&$\mathrm{	-69	^{\circ }	16	\arcmin	}$&	72 	&$\mathrm{	4	h	49.5	m	}$&$\mathrm{	-69	^{\circ }	14	\arcmin	}$&	74 	&	2.7 	&	91 	&		\\
5 &LMC N J0449-6826	&II	&$\mathrm{	4	h	49.5	m	}$&$\mathrm{	-68	^{\circ }	28	\arcmin	}$&	100 	&$\mathrm{	4	h	50.1	m	}$&$\mathrm{	-68	^{\circ }	32	\arcmin	}$&	77 	&	1.7 	&	152 	&		\\
9 &LMC N J0450-6930	&II	&$\mathrm{	4	h	50.5	m	}$&$\mathrm{	-69	^{\circ }	36	\arcmin	}$&	33 	&$\mathrm{	4	h	50.2	m	}$&$\mathrm{	-69	^{\circ }	34	\arcmin	}$&	55 	&	2.1 	&	83 	&		\\
11 &LMC N J0451-6858	&I	&$\mathrm{	4	h	51.3	m	}$&$\mathrm{	-69	^{\circ }	4	\arcmin	}$&	29 	&$\mathrm{	4	h	51.3	m	}$&$\mathrm{	-69	^{\circ }	4	\arcmin	}$&	29 	&	2.7 	&	0 	&		\\
12 &LMC N J0451-6704	&III	&$\mathrm{	4	h	51.9	m	}$&$\mathrm{	-67	^{\circ }	8	\arcmin	}$&	108 	&$\mathrm{	4	h	52.2	m	}$&$\mathrm{	-67	^{\circ }	8	\arcmin	}$&	62 	&	3.2 	&	70 	&		\\
17 &LMC N J0453-6909	&III	&$\mathrm{	4	h	54.1	m	}$&$\mathrm{	-69	^{\circ }	14	\arcmin	}$&	85 	&$\mathrm{	4	h	53.3	m	}$&$\mathrm{	-69	^{\circ }	12	\arcmin	}$&	59 	&	2.1 	&	159 	&		\\
22 &LMC N J0455-6830	&II	&$\mathrm{	4	h	55.9	m	}$&$\mathrm{	-68	^{\circ }	36	\arcmin	}$&	47 	&$\mathrm{	4	h	55.9	m	}$&$\mathrm{	-68	^{\circ }	36	\arcmin	}$&	18 	&	2.2 	&	0 	&		\\
23 &LMC N J0455-6634	&III	&$\mathrm{	4	h	56.4	m	}$&$\mathrm{	-66	^{\circ }	42	\arcmin	}$&	64 	&$\mathrm{	4	h	56.4	m	}$&$\mathrm{	-66	^{\circ }	40	\arcmin	}$&	*	&	1.9 	&	28 	&	30,32,33,236	\\
24 &LMC N J0455-6930	&III	&$\mathrm{	4	h	56.3	m	}$&$\mathrm{	-69	^{\circ }	36	\arcmin	}$&	29 	&$\mathrm{	4	h	56.3	m	}$&$\mathrm{	-69	^{\circ }	36	\arcmin	}$&	18 	&	0.8 	&	0 	&		\\
26 &LMC N J0457-6844	&III	&$\mathrm{	4	h	57.5	m	}$&$\mathrm{	-68	^{\circ }	48	\arcmin	}$&	29 	&$\mathrm{	4	h	57.5	m	}$&$\mathrm{	-68	^{\circ }	48	\arcmin	}$&	24 	&	1.5 	&	0 	&		\\
27 &LMC N J0457-6826	&III	&$\mathrm{	4	h	57.4	m	}$&$\mathrm{	-68	^{\circ }	30	\arcmin	}$&	40 	&$\mathrm{	4	h	57.1	m	}$&$\mathrm{	-68	^{\circ }	32	\arcmin	}$&	18 	&	1.1 	&	91 	&		\\
35 &LMC N J0459-6614	&II	&$\mathrm{	4	h	58.8	m	}$&$\mathrm{	-66	^{\circ }	18	\arcmin	}$&	79 	&$\mathrm{	4	h	58.4	m	}$&$\mathrm{	-66	^{\circ }	24	\arcmin	}$&	85 	&	2.7 	&	130 	&	30,32,33,236	\\
36 &LMC N J0500-6622	&II	&$\mathrm{	5	h	0.7	m	}$&$\mathrm{	-66	^{\circ }	26	\arcmin	}$&	52 	&$\mathrm{	5	h	0.7	m	}$&$\mathrm{	-66	^{\circ }	26	\arcmin	}$&	37 	&	1.5 	&	0 	&		\\
38 &LMC N J0502-6903	&II	&$\mathrm{	5	h	2.4	m	}$&$\mathrm{	-69	^{\circ }	6	\arcmin	}$&	37 	&$\mathrm{	5	h	2.4	m	}$&$\mathrm{	-69	^{\circ }	6	\arcmin	}$&	41 	&	2.1 	&	0 	&		\\
39 &LMC N J0503-6553	&II	&$\mathrm{	5	h	3	m	}$&$\mathrm{	-65	^{\circ }	58	\arcmin	}$&	40 	&$\mathrm{	5	h	3	m	}$&$\mathrm{	-65	^{\circ }	58	\arcmin	}$&	62 	&	2.7 	&	0 	&		\\
40 &LMC N J0503-6540	&II	&$\mathrm{	5	h	3.1	m	}$&$\mathrm{	-65	^{\circ }	44	\arcmin	}$&	37 	&$\mathrm{	5	h	3.4	m	}$&$\mathrm{	-65	^{\circ }	46	\arcmin	}$&	24 	&	0.6 	&	75 	&		\\
41 &LMC N J0503-6828	&II	&$\mathrm{	5	h	3.6	m	}$&$\mathrm{	-68	^{\circ }	32	\arcmin	}$&	33 	&$\mathrm{	5	h	4.3	m	}$&$\mathrm{	-68	^{\circ }	36	\arcmin	}$&	37 	&	1.1 	&	160 	&		\\
43 &LMC N J0503-6643	&II	&$\mathrm{	5	h	3.4	m	}$&$\mathrm{	-66	^{\circ }	48	\arcmin	}$&	33 	&$\mathrm{	5	h	3.4	m	}$&$\mathrm{	-66	^{\circ }	50	\arcmin	}$&	24 	&	1.2 	&	26 	&		\\
45 &LMC N J0503-6719	&III	&$\mathrm{	5	h	3.7	m	}$&$\mathrm{	-67	^{\circ }	24	\arcmin	}$&	44 	&$\mathrm{	5	h	3.7	m	}$&$\mathrm{	-67	^{\circ }	24	\arcmin	}$&	24 	&	2.0 	&	0 	&		\\
46 &LMC N J0504-6802	&III	&$\mathrm{	5	h	4.6	m	}$&$\mathrm{	-68	^{\circ }	6	\arcmin	}$&	47 	&$\mathrm{	5	h	4.3	m	}$&$\mathrm{	-68	^{\circ }	8	\arcmin	}$&	18 	&	1.0 	&	83 	&		\\
48 &LMC N J0504-7007	&II	&$\mathrm{	5	h	5	m	}$&$\mathrm{	-70	^{\circ }	12	\arcmin	}$&	44 	&$\mathrm{	5	h	5.1	m	}$&$\mathrm{	-70	^{\circ }	12	\arcmin	}$&	8 	&	1.4 	&	0 	&		\\
49 &LMC N J0504-7056	&III	&$\mathrm{	5	h	5.3	m	}$&$\mathrm{	-71	^{\circ }	0	\arcmin	}$&	52 	&$\mathrm{	5	h	5.3	m	}$&$\mathrm{	-71	^{\circ }	0	\arcmin	}$&	34 	&	1.0 	&	0 	&		\\
52 &LMC N J0506-6753	&II	&$\mathrm{	5	h	6.8	m	}$&$\mathrm{	-67	^{\circ }	58	\arcmin	}$&	23 	&$\mathrm{	5	h	7.2	m	}$&$\mathrm{	-67	^{\circ }	56	\arcmin	}$&	24 	&	1.0 	&	83 	&		\\
53 &LMC N J0507-7041	&II	&$\mathrm{	5	h	7.9	m	}$&$\mathrm{	-70	^{\circ }	46	\arcmin	}$&	23 	&$\mathrm{	5	h	7.8	m	}$&$\mathrm{	-70	^{\circ }	48	\arcmin	}$&	24 	&	0.9 	&	28 	&		\\
55 &LMC N J0507-6858	&I	&$\mathrm{	5	h	8.1	m	}$&$\mathrm{	-69	^{\circ }	2	\arcmin	}$&	47 	&$\mathrm{	5	h	8.1	m	}$&$\mathrm{	-69	^{\circ }	0	\arcmin	}$&	29 	&	1.3 	&	28 	&		\\
57 &LMC N J0508-6905	&I	&$\mathrm{	5	h	9.1	m	}$&$\mathrm{	-69	^{\circ }	8	\arcmin	}$&	23 	&$\mathrm{	5	h	9.2	m	}$&$\mathrm{	-60	^{\circ }	8	\arcmin	}$&	8 	&	1.0 	&	0 	&		\\
60 &LMC N J0509-6827	&II	&$\mathrm{	5	h	9.4	m	}$&$\mathrm{	-68	^{\circ }	32	\arcmin	}$&	33 	&$\mathrm{	5	h	9.5	m	}$&$\mathrm{	-68	^{\circ }	30	\arcmin	}$&	47 	&	1.6 	&	26 	&		\\
61 &LMC N J0509-7049	&II	&$\mathrm{	5	h	10.2	m	}$&$\mathrm{	-70	^{\circ }	54	\arcmin	}$&	29 	&$\mathrm{	5	h	9.8	m	}$&$\mathrm{	-70	^{\circ }	54	\arcmin	}$&	8 	&	1.0 	&	87 	&		\\
62 &LMC N J0509-6912	&I	&$\mathrm{	5	h	10.2	m	}$&$\mathrm{	-69	^{\circ }	16	\arcmin	}$&	52 	&$\mathrm{	5	h	10.6	m	}$&$\mathrm{	-69	^{\circ }	16	\arcmin	}$&	37 	&	1.7 	&	79 	&		\\
63 &LMC N J0510-6853	&III	&$\mathrm{	5	h	10.4	m	}$&$\mathrm{	-68	^{\circ }	56	\arcmin	}$&	84 	&$\mathrm{	5	h	10	m	}$&$\mathrm{	-68	^{\circ }	52	\arcmin	}$&	77 	&	2.3 	&	93 	&	245,68	\\
64 &LMC N J0510-6706	&II	&$\mathrm{	5	h	10.7	m	}$&$\mathrm{	-67	^{\circ }	12	\arcmin	}$&	52 	&$\mathrm{	5	h	11.1	m	}$&$\mathrm{	-66	^{\circ }	58	\arcmin	}$&	74 	&	1.2 	&	222 	&	67	\\
65 &LMC N J0511-6927	&II	&$\mathrm{	5	h	11.6	m	}$&$\mathrm{	-69	^{\circ }	30	\arcmin	}$&	23 	&$\mathrm{	5	h	12	m	}$&$\mathrm{	-69	^{\circ }	32	\arcmin	}$&	47 	&	0.8 	&	83 	&	244,71	\\
67 &LMC N J0512-6710	&II	&$\mathrm{	5	h	12.1	m	}$&$\mathrm{	-67	^{\circ }	14	\arcmin	}$&	55 	&$\mathrm{	5	h	12.4	m	}$&$\mathrm{	-67	^{\circ }	14	\arcmin	}$&	55 	&	1.6 	&	79 	&	64	\\
68 &LMC N J0512-6903	&II	&$\mathrm{	5	h	12.2	m	}$&$\mathrm{	-69	^{\circ }	4	\arcmin	}$&	80 	&$\mathrm{	5	h	11.4	m	}$&$\mathrm{	-69	^{\circ }	4	\arcmin	}$&	88 	&	1.8 	&	166 	&	63	\\
69 &LMC N J0512-7028	&II	&$\mathrm{	5	h	12.8	m	}$&$\mathrm{	-70	^{\circ }	32	\arcmin	}$&	68 	&$\mathrm{	5	h	12.8	m	}$&$\mathrm{	-70	^{\circ }	30	\arcmin	}$&	50 	&	1.5 	&	26 	&		\\
71 &LMC N J0513-6936	&II	&$\mathrm{	5	h	13.5	m	}$&$\mathrm{	-69	^{\circ }	40	\arcmin	}$&	79 	&$\mathrm{	5	h	13.5	m	}$&$\mathrm{	-69	^{\circ }	44	\arcmin	}$&	59 	&	1.9 	&	62 	&		\\
72 &LMC N J0513-6922	&III	&$\mathrm{	5	h	13.5	m	}$&$\mathrm{	-69	^{\circ }	26	\arcmin	}$&	49 	&$\mathrm{	5	h	13.2	m	}$&$\mathrm{	-69	^{\circ }	24	\arcmin	}$&	34 	&	1.7 	&	83 	&		\\
74 &LMC N J0514-7010	&II	&$\mathrm{	5	h	14.9	m	}$&$\mathrm{	-70	^{\circ }	14	\arcmin	}$&	66 	&$\mathrm{	5	h	14.9	m	}$&$\mathrm{	-70	^{\circ }	14	\arcmin	}$&	34 	&	1.6 	&	0 	&		\\
77 &LMC N J0515-7034	&I	&$\mathrm{	5	h	15.6	m	}$&$\mathrm{	-70	^{\circ }	38	\arcmin	}$&	44 	&$\mathrm{	5	h	15.6	m	}$&$\mathrm{	-70	^{\circ }	38	\arcmin	}$&	18 	&	0.6 	&	0 	&		\\
80 &LMC N J0516-6807	&II	&$\mathrm{	5	h	16.1	m	}$&$\mathrm{	-68	^{\circ }	2	\arcmin	}$&	124 	&$\mathrm{	5	h	17.2	m	}$&$\mathrm{	-68	^{\circ }	4	\arcmin	}$&	62 	&	1.4 	&	237 	&		\\
81 &LMC N J0516-6616	&I	&$\mathrm{	5	h	16	m	}$&$\mathrm{	-66	^{\circ }	20	\arcmin	}$&	29 	&$\mathrm{	5	h	16	m	}$&$\mathrm{	-66	^{\circ }	18	\arcmin	}$&	24 	&	0.3 	&	26 	&		\\
83 &LMC N J0516-6922	&II	&$\mathrm{	5	h	17	m	}$&$\mathrm{	-69	^{\circ }	26	\arcmin	}$&	23 	&$\mathrm{	5	h	17.4	m	}$&$\mathrm{	-69	^{\circ }	26	\arcmin	}$&	34 	&	1.3 	&	87 	&	91	\\
84 &LMC N J0516-6559	&II	&$\mathrm{	5	h	16.7	m	}$&$\mathrm{	-66	^{\circ }	2	\arcmin	}$&	29 	&$\mathrm{	5	h	17.1	m	}$&$\mathrm{	-66	^{\circ }	2	\arcmin	}$&	29 	&	0.6 	&	70 	&		\\
86 &LMC N J0517-7114	&III	&$\mathrm{	5	h	18.3	m	}$&$\mathrm{	-71	^{\circ }	18	\arcmin	}$&	44 	&$\mathrm{	5	h	18.3	m	}$&$\mathrm{	-71	^{\circ }	20	\arcmin	}$&	37 	&	1.3 	&	26 	&		\\
89 &LMC N J0517-6642	&II	&$\mathrm{	5	h	17.6	m	}$&$\mathrm{	-66	^{\circ }	44	\arcmin	}$&	40 	&$\mathrm{	5	h	17.6	m	}$&$\mathrm{	-66	^{\circ }	44	\arcmin	}$&	18 	&	0.9 	&	0 	&		\\
90 &LMC N J0517-6932	&II	&$\mathrm{	5	h	18.5	m	}$&$\mathrm{	-69	^{\circ }	36	\arcmin	}$&	33 	&$\mathrm{	5	h	18.5	m	}$&$\mathrm{	-69	^{\circ }	36	\arcmin	}$&	*	&	0.9 	&	0 	&		\\
92 &LMC N J0518-7001	&I	&$\mathrm{	5	h	18.8	m	}$&$\mathrm{	-70	^{\circ }	6	\arcmin	}$&	33 	&$\mathrm{	5	h	18.1	m	}$&$\mathrm{	-70	^{\circ }	4	\arcmin	}$&	18 	&	1.1 	&	168 	&		\\
93 &LMC N J0518-6620	&I	&$\mathrm{	5	h	18.3	m	}$&$\mathrm{	-66	^{\circ }	24	\arcmin	}$&	23 	&$\mathrm{	5	h	18	m	}$&$\mathrm{	-66	^{\circ }	22	\arcmin	}$&	*	&	0.5 	&	75 	&		\\
94 &LMC N J0518-6951	&II	&$\mathrm{	5	h	19.2	m	}$&$\mathrm{	-69	^{\circ }	54	\arcmin	}$&	23 	&$\mathrm{	5	h	19.2	m	}$&$\mathrm{	-69	^{\circ }	56	\arcmin	}$&	24 	&	0.9 	&	26 	&		\\
95 &LMC N J0519-6625	&I	&$\mathrm{	5	h	19	m	}$&$\mathrm{	-66	^{\circ }	28	\arcmin	}$&	23 	&$\mathrm{	5	h	19	m	}$&$\mathrm{	-66	^{\circ }	30	\arcmin	}$&	*	&	1.2 	&	26 	&	100	\\
96 &LMC N J0519-6938	&III	&$\mathrm{	5	h	19.6	m	}$&$\mathrm{	-69	^{\circ }	40	\arcmin	}$&	72 	&$\mathrm{	5	h	19.6	m	}$&$\mathrm{	-69	^{\circ }	40	\arcmin	}$&	29 	&	2.0 	&	0 	&		\\
97 &LMC N J0519-7113	&I	&$\mathrm{	5	h	20	m	}$&$\mathrm{	-71	^{\circ }	18	\arcmin	}$&	40 	&$\mathrm{	5	h	20.4	m	}$&$\mathrm{	-71	^{\circ }	16	\arcmin	}$&	24 	&	0.9 	&	91 	&		\\
99 &LMC N J0520-7043	&I	&$\mathrm{	5	h	20.4	m	}$&$\mathrm{	-70	^{\circ }	44	\arcmin	}$&	52 	&$\mathrm{	5	h	20	m	}$&$\mathrm{	-70	^{\circ }	46	\arcmin	}$&	37 	&	1.0 	&	91 	&		\\
100 &LMC N J0520-665	&II	&$\mathrm{	5	h	20	m	}$&$\mathrm{	-66	^{\circ }	54	\arcmin	}$&	37 	&$\mathrm{	5	h	21	m	}$&$\mathrm{	-66	^{\circ }	50	\arcmin	}$&	52 	&	1.3 	&	226 	&		\\
103 &LMC N J0521-714	&II	&$\mathrm{	5	h	21.7	m	}$&$\mathrm{	-71	^{\circ }	46	\arcmin	}$&	23 	&$\mathrm{	5	h	21.7	m	}$&$\mathrm{	-71	^{\circ }	46	\arcmin	}$&	24 	&	1.2 	&	0 	&		\\
104 &LMC N J0521-701	&II	&$\mathrm{	5	h	21.6	m	}$&$\mathrm{	-70	^{\circ }	16	\arcmin	}$&	49 	&$\mathrm{	5	h	21.2	m	}$&$\mathrm{	-70	^{\circ }	18	\arcmin	}$&	29 	&	0.8 	&	91 	&		\\
105 &LMC N J0521-700	&II	&$\mathrm{	5	h	21.6	m	}$&$\mathrm{	-70	^{\circ }	4	\arcmin	}$&	79 	&$\mathrm{	5	h	21.6	m	}$&$\mathrm{	-70	^{\circ }	6	\arcmin	}$&	44 	&	1.4 	&	26 	&		\\
106 &LMC N J0521-684	&II	&$\mathrm{	5	h	21.5	m	}$&$\mathrm{	-68	^{\circ }	44	\arcmin	}$&	29 	&$\mathrm{	5	h	21.5	m	}$&$\mathrm{	-68	^{\circ }	44	\arcmin	}$&	8 	&	1.7 	&	0 	&		\\
108 &LMC N J0521-684	&II	&$\mathrm{	5	h	21.8	m	}$&$\mathrm{	-68	^{\circ }	50	\arcmin	}$&	55 	&$\mathrm{	5	h	21.8	m	}$&$\mathrm{	-68	^{\circ }	50	\arcmin	}$&	34 	&	1.4 	&	0 	&		\\
110 &LMC N J0522-694	&III	&$\mathrm{	5	h	22.7	m	}$&$\mathrm{	-69	^{\circ }	44	\arcmin	}$&	52 	&$\mathrm{	5	h	22.3	m	}$&$\mathrm{	-69	^{\circ }	44	\arcmin	}$&	18 	&	1.2 	&	79 	&		\\
115 &LMC N J0522-654	&III	&$\mathrm{	5	h	22.6	m	}$&$\mathrm{	-65	^{\circ }	44	\arcmin	}$&	57 	&$\mathrm{	5	h	22.6	m	}$&$\mathrm{	-65	^{\circ }	42	\arcmin	}$&	44 	&	1.7 	&	26 	&		\\
118 &LMC N J0523-682	&I	&$\mathrm{	5	h	22.9	m	}$&$\mathrm{	-68	^{\circ }	24	\arcmin	}$&	44 	&$\mathrm{	5	h	22.9	m	}$&$\mathrm{	-68	^{\circ }	26	\arcmin	}$&	37 	&	3.2 	&	26 	&		\\
119 &LMC N J0523-664	&II	&$\mathrm{	5	h	23	m	}$&$\mathrm{	-66	^{\circ }	46	\arcmin	}$&	57 	&$\mathrm{	5	h	22.7	m	}$&$\mathrm{	-66	^{\circ }	46	\arcmin	}$&	41 	&	2.5 	&	79 	&		\\
123 &LMC N J0523-713	&II	&$\mathrm{	5	h	24.2	m	}$&$\mathrm{	-71	^{\circ }	42	\arcmin	}$&	49 	&$\mathrm{	5	h	26.4	m	}$&$\mathrm{	-71	^{\circ }	38	\arcmin	}$&	62 	&	1.4 	&	466 	&	131,249	\\
126 &LMC N J0524-672	&II	&$\mathrm{	5	h	24.5	m	}$&$\mathrm{	-67	^{\circ }	30	\arcmin	}$&	44 	&$\mathrm{	5	h	24.5	m	}$&$\mathrm{	-67	^{\circ }	30	\arcmin	}$&	24 	&	2.3 	&	0 	&		\\
127 &LMC N J0524-702	&I	&$\mathrm{	5	h	24.8	m	}$&$\mathrm{	-70	^{\circ }	30	\arcmin	}$&	37 	&$\mathrm{	5	h	24.4	m	}$&$\mathrm{	-70	^{\circ }	30	\arcmin	}$&	24 	&	0.6 	&	87 	&		\\
130 &LMC N J0524-691	&II	&$\mathrm{	5	h	24.9	m	}$&$\mathrm{	-69	^{\circ }	18	\arcmin	}$&	33 	&$\mathrm{	5	h	24.5	m	}$&$\mathrm{	-69	^{\circ }	12	\arcmin	}$&	47 	&	1.2 	&	123 	&		\\
131 &LMC N J0524-713	&II	&$\mathrm{	5	h	25.5	m	}$&$\mathrm{	-71	^{\circ }	36	\arcmin	}$&	29 	&$\mathrm{	5	h	25.9	m	}$&$\mathrm{	-71	^{\circ }	38	\arcmin	}$&	64 	&	1.7 	&	99 	&	123,249	\\
132 &LMC N J0525-694	&II	&$\mathrm{	5	h	26.2	m	}$&$\mathrm{	-69	^{\circ }	52	\arcmin	}$&	33 	&$\mathrm{	5	h	25.8	m	}$&$\mathrm{	-69	^{\circ }	56	\arcmin	}$&	34 	&	1.0 	&	99 	&		\\
133 &LMC N J0525-691	&III	&$\mathrm{	5	h	26	m	}$&$\mathrm{	-69	^{\circ }	20	\arcmin	}$&	33 	&$\mathrm{	5	h	24.5	m	}$&$\mathrm{	-69	^{\circ }	12	\arcmin	}$&	47 	&	0.8 	&	359 	&	130	\\
134 &LMC N J0525-662	&II	&$\mathrm{	5	h	25.7	m	}$&$\mathrm{	-66	^{\circ }	24	\arcmin	}$&	23 	&$\mathrm{	5	h	25.3	m	}$&$\mathrm{	-66	^{\circ }	26	\arcmin	}$&	116 	&	3.8 	&	75 	&	135	\\
136 &LMC N J0526-683	&II	&$\mathrm{	5	h	26.2	m	}$&$\mathrm{	-68	^{\circ }	38	\arcmin	}$&	37 	&$\mathrm{	5	h	26.2	m	}$&$\mathrm{	-68	^{\circ }	38	\arcmin	}$&	41 	&	1.4 	&	0 	&		\\
139 &LMC N J0526-684	&III	&$\mathrm{	5	h	26.6	m	}$&$\mathrm{	-68	^{\circ }	50	\arcmin	}$&	29 	&$\mathrm{	5	h	26.3	m	}$&$\mathrm{	-68	^{\circ }	48	\arcmin	}$&	29 	&	1.9 	&	91 	&		\\
141 &LMC N J0526-655	&II	&$\mathrm{	5	h	26.5	m	}$&$\mathrm{	-65	^{\circ }	56	\arcmin	}$&	23 	&$\mathrm{	5	h	26.2	m	}$&$\mathrm{	-66	^{\circ }	0	\arcmin	}$&	160 	&	2.0 	&	93 	&	134,135,250	\\
143 &LMC N J0526-711	&II	&$\mathrm{	5	h	27.5	m	}$&$\mathrm{	-71	^{\circ }	22	\arcmin	}$&	97 	&$\mathrm{	5	h	26	m	}$&$\mathrm{	-71	^{\circ }	26	\arcmin	}$&	120 	&	1.3 	&	346 	&	153,156	\\
145 &LMC N J0527-703	&II	&$\mathrm{	5	h	28	m	}$&$\mathrm{	-70	^{\circ }	38	\arcmin	}$&	40 	&$\mathrm{	5	h	28.1	m	}$&$\mathrm{	-70	^{\circ }	40	\arcmin	}$&	18 	&	0.6 	&	26 	&		\\
146 &LMC N J0527-705	&II	&$\mathrm{	5	h	28.6	m	}$&$\mathrm{	-70	^{\circ }	54	\arcmin	}$&	37 	&$\mathrm{	5	h	28.6	m	}$&$\mathrm{	-70	^{\circ }	54	\arcmin	}$&	72 	&	1.0 	&	0 	&	149	\\
149 &LMC N J0528-705	&II	&$\mathrm{	5	h	29	m	}$&$\mathrm{	-71	^{\circ }	0	\arcmin	}$&	33 	&$\mathrm{	5	h	29.4	m	}$&$\mathrm{	-71	^{\circ }	0	\arcmin	}$&	24 	&	1.7 	&	87 	&		\\
150 &LMC N J0529-683	&II	&$\mathrm{	5	h	30.2	m	}$&$\mathrm{	-68	^{\circ }	32	\arcmin	}$&	29 	&$\mathrm{	5	h	30.6	m	}$&$\mathrm{	-68	^{\circ }	36	\arcmin	}$&	142 	&	1.8 	&	106 	&	154,163	\\
153 &LMC N J0530-710	&III	&$\mathrm{	5	h	31.6	m	}$&$\mathrm{	-71	^{\circ }	10	\arcmin	}$&	80 	&$\mathrm{	5	h	31.6	m	}$&$\mathrm{	-71	^{\circ }	10	\arcmin	}$&	47 	&	2.6 	&	0 	&		\\
154 &LMC N J0531-683	&III	&$\mathrm{	5	h	31.7	m	}$&$\mathrm{	-68	^{\circ }	34	\arcmin	}$&	108 	&$\mathrm{	5	h	32.7	m	}$&$\mathrm{	-68	^{\circ }	28	\arcmin	}$&	55 	&	2.7 	&	243 	&	150,163	\\
155 &LMC N J0532-674	&III	&$\mathrm{	5	h	32	m	}$&$\mathrm{	-67	^{\circ }	46	\arcmin	}$&	68 	&$\mathrm{	5	h	32.7	m	}$&$\mathrm{	-67	^{\circ }	46	\arcmin	}$&	44 	&	3.1 	&	157 	&		\\
156 &LMC N J0532-711	&II	&$\mathrm{	5	h	32.9	m	}$&$\mathrm{	-71	^{\circ }	16	\arcmin	}$&	47 	&$\mathrm{	5	h	32.9	m	}$&$\mathrm{	-71	^{\circ }	16	\arcmin	}$&	41 	&	2.5 	&	0 	&		\\
157 &LMC N J0532-683	&II	&$\mathrm{	5	h	32.5	m	}$&$\mathrm{	-68	^{\circ }	40	\arcmin	}$&	40 	&$\mathrm{	5	h	32.5	m	}$&$\mathrm{	-68	^{\circ }	40	\arcmin	}$&	41 	&	2.7 	&	0 	&		\\
158 &LMC N J0532-662	&III	&$\mathrm{	5	h	32.4	m	}$&$\mathrm{	-66	^{\circ }	30	\arcmin	}$&	47 	&$\mathrm{	5	h	32	m	}$&$\mathrm{	-66	^{\circ }	26	\arcmin	}$&	29 	&	0.9 	&	99 	&		\\
162 &LMC N J0532-685	&II	&$\mathrm{	5	h	33.4	m	}$&$\mathrm{	-69	^{\circ }	0	\arcmin	}$&	33 	&$\mathrm{	5	h	33	m	}$&$\mathrm{	-69	^{\circ }	0	\arcmin	}$&	18 	&	2.4 	&	79 	&		\\
171 &LMC N J0535-690	&III	&$\mathrm{	5	h	36	m	}$&$\mathrm{	-69	^{\circ }	4	\arcmin	}$&	66 	&$\mathrm{	5	h	36.4	m	}$&$\mathrm{	-69	^{\circ }	4	\arcmin	}$&	91 	&	3.5 	&	79 	&	183	\\
172 &LMC N J0535-684	&II	&$\mathrm{	5	h	36.2	m	}$&$\mathrm{	-68	^{\circ }	46	\arcmin	}$&	37 	&$\mathrm{	5	h	35.8	m	}$&$\mathrm{	-68	^{\circ }	46	\arcmin	}$&	41 	&	1.6 	&	79 	&		\\
179 &LMC N J0537-661	&III	&$\mathrm{	5	h	37.3	m	}$&$\mathrm{	-66	^{\circ }	20	\arcmin	}$&	47 	&$\mathrm{	5	h	37.3	m	}$&$\mathrm{	-66	^{\circ }	20	\arcmin	}$&	44 	&	2.1 	&	0 	&		\\
180 &LMC N J0537-662	&II	&$\mathrm{	5	h	37.4	m	}$&$\mathrm{	-66	^{\circ }	28	\arcmin	}$&	23 	&$\mathrm{	5	h	37.1	m	}$&$\mathrm{	-66	^{\circ }	30	\arcmin	}$&	34 	&	1.8 	&	75 	&		\\
184 &LMC N J0538-693	&II	&$\mathrm{	5	h	38.7	m	}$&$\mathrm{	-69	^{\circ }	36	\arcmin	}$&	33 	&$\mathrm{	5	h	39.1	m	}$&$\mathrm{	-69	^{\circ }	34	\arcmin	}$&	41 	&	3.5 	&	83 	&	191	\\
190 &LMC N J0538-685	&III	&$\mathrm{	5	h	38.9	m	}$&$\mathrm{	-68	^{\circ }	56	\arcmin	}$&	23 	&$\mathrm{	5	h	41.1	m	}$&$\mathrm{	-68	^{\circ }	54	\arcmin	}$&	81 	&	2.9 	&	480 	&		\\
191 &LMC N J0539-693	&III	&$\mathrm{	5	h	39.4	m	}$&$\mathrm{	-69	^{\circ }	32	\arcmin	}$&	57 	&$\mathrm{	5	h	39.1	m	}$&$\mathrm{	-69	^{\circ }	34	\arcmin	}$&	41 	&	3.5 	&	83 	&	184	\\
205 &LMC N J0542-711	&III	&$\mathrm{	5	h	42.1	m	}$&$\mathrm{	-71	^{\circ }	20	\arcmin	}$&	64 	&$\mathrm{	5	h	42	m	}$&$\mathrm{	-71	^{\circ }	18	\arcmin	}$&	47 	&	2.5 	&	28 	&		\\
207 &LMC N J0542-694	&II	&$\mathrm{	5	h	43.5	m	}$&$\mathrm{	-69	^{\circ }	46	\arcmin	}$&	23 	&$\mathrm{	5	h	43.5	m	}$&$\mathrm{	-69	^{\circ }	46	\arcmin	}$&	18 	&	4.1 	&	0 	&	266	\\
208 &LMC N J0543-675	&III	&$\mathrm{	5	h	43.5	m	}$&$\mathrm{	-67	^{\circ }	58	\arcmin	}$&	37 	&$\mathrm{	5	h	43.8	m	}$&$\mathrm{	-67	^{\circ }	58	\arcmin	}$&	34 	&	2.0 	&	79 	&	211	\\
209 &LMC N J0543-692	&II	&$\mathrm{	5	h	43.9	m	}$&$\mathrm{	-69	^{\circ }	28	\arcmin	}$&	33 	&$\mathrm{	5	h	44.2	m	}$&$\mathrm{	-69	^{\circ }	22	\arcmin	}$&	47 	&	3.0 	&	106 	&	214,216,268	\\
211 &LMC N J0543-675	&III	&$\mathrm{	5	h	43.8	m	}$&$\mathrm{	-67	^{\circ }	56	\arcmin	}$&	23 	&$\mathrm{	5	h	43.8	m	}$&$\mathrm{	-67	^{\circ }	58	\arcmin	}$&	29 	&	1.8 	&	28 	&	208	\\
214 &LMC N J0543-691	&II	&$\mathrm{	5	h	43.4	m	}$&$\mathrm{	-69	^{\circ }	16	\arcmin	}$&	33 	&$\mathrm{	5	h	44.2	m	}$&$\mathrm{	-69	^{\circ }	22	\arcmin	}$&	44 	&	2.5 	&	211 	&	209,216,268	\\
215 &LMC N J0544-712	&I	&$\mathrm{	5	h	45.2	m	}$&$\mathrm{	-71	^{\circ }	28	\arcmin	}$&	59 	&$\mathrm{	5	h	44.7	m	}$&$\mathrm{	-71	^{\circ }	28	\arcmin	}$&	57 	&	1.9 	&	87 	&		\\
218 &LMC N J0544-671	&II	&$\mathrm{	5	h	44.9	m	}$&$\mathrm{	-67	^{\circ }	20	\arcmin	}$&	29 	&$\mathrm{	5	h	44.4	m	}$&$\mathrm{	-67	^{\circ }	28	\arcmin	}$&	50 	&	1.0 	&	167 	&	213	\\
220 &LMC N J0545-694	&III	&$\mathrm{	5	h	45.9	m	}$&$\mathrm{	-69	^{\circ }	50	\arcmin	}$&	47 	&$\mathrm{	5	h	45.6	m	}$&$\mathrm{	-69	^{\circ }	52	\arcmin	}$&	37 	&	2.0 	&	83 	&		\\
222 &LMC N J0546-710	&I	&$\mathrm{	5	h	46.8	m	}$&$\mathrm{	-71	^{\circ }	6	\arcmin	}$&	44 	&$\mathrm{	5	h	46.8	m	}$&$\mathrm{	-71	^{\circ }	6	\arcmin	}$&	44 	&	1.7 	&	0 	&		\\
223 &LMC N J0546-693	&II	&$\mathrm{	5	h	46.8	m	}$&$\mathrm{	-69	^{\circ }	38	\arcmin	}$&	52 	&$\mathrm{	5	h	47.2	m	}$&$\mathrm{	-69	^{\circ }	40	\arcmin	}$&	57 	&	3.3 	&	99 	&		\\
224 &LMC N J0547-680	&I	&$\mathrm{	5	h	47.6	m	}$&$\mathrm{	-68	^{\circ }	10	\arcmin	}$&	33 	&$\mathrm{	5	h	48	m	}$&$\mathrm{	-68	^{\circ }	10	\arcmin	}$&	55 	&	1.7 	&	79 	&		\\
225 &LMC N J0547-704	&I	&$\mathrm{	5	h	47.8	m	}$&$\mathrm{	-70	^{\circ }	42	\arcmin	}$&	87 	&$\mathrm{	5	h	47.8	m	}$&$\mathrm{	-70	^{\circ }	40	\arcmin	}$&	*	&	1.3 	&	28 	&	228	\\
226 &LMC N J0547-6953	&II	&$\mathrm{	5	h	48.7	m	}$&$\mathrm{	-69	^{\circ }	54	\arcmin	}$&	61 	&$\mathrm{	5	h	48	m	}$&$\mathrm{	-69	^{\circ }	56	\arcmin	}$&	104 	&	1.6 	&	159 	&	223,227,270	\\
230 &LMC N J0555-681	&III	&$\mathrm{	5	h	55.9	m	}$&$\mathrm{	-68	^{\circ }	10	\arcmin	}$&	73 	&$\mathrm{	5	h	55.9	m	}$&$\mathrm{	-68	^{\circ }	10	\arcmin	}$&	*	&	1.3 	&	0 	&	271	\\
231 &LMC N J0447-672	&I	&$\mathrm{	4	h	47.2	m	}$&$\mathrm{	-67	^{\circ }	34	\arcmin	}$&	17 	&$\mathrm{	4	h	47.1	m	}$&$\mathrm{	-67	^{\circ }	36	\arcmin	}$&	190 	&	1.2 	&	28 	&	2,232	\\
232 &LMC N J0448-672	&I	&$\mathrm{	4	h	48	m	}$&$\mathrm{	-67	^{\circ }	28	\arcmin	}$&	17 	&$\mathrm{	4	h	47.2	m	}$&$\mathrm{	-67	^{\circ }	18	\arcmin	}$&	82 	&	1.2 	&	229 	&	2	\\
233 &LMC N J0449-681	&II	&$\mathrm{	4	h	49.8	m	}$&$\mathrm{	-68	^{\circ }	16	\arcmin	}$&	17 	&$\mathrm{	4	h	49.5	m	}$&$\mathrm{	-68	^{\circ }	12	\arcmin	}$&	88 	&	1.3 	&	86 	&	6	\\
236 &LMC N J0458-661	&II	&$\mathrm{	4	h	58.7	m	}$&$\mathrm{	-66	^{\circ }	22	\arcmin	}$&	17 	&$\mathrm{	4	h	58.4	m	}$&$\mathrm{	-66	^{\circ }	24	\arcmin	}$&	75 	&	3.7 	&	83 	&	30,32,33,35	\\
237 &LMC N J0459-660	&II	&$\mathrm{	4	h	59.8	m	}$&$\mathrm{	-66	^{\circ }	12	\arcmin	}$&	17 	&$\mathrm{	5	h	0.2	m	}$&$\mathrm{	-66	^{\circ }	12	\arcmin	}$&	*	&	2.7 	&	70 	&		\\
243 &LMC N J0511-705	&I	&$\mathrm{	5	h	12.3	m	}$&$\mathrm{	-70	^{\circ }	54	\arcmin	}$&	17 	&$\mathrm{	5	h	12.7	m	}$&$\mathrm{	-70	^{\circ }	54	\arcmin	}$&	8 	&	0.7 	&	87 	&		\\
244 &LMC N J0511-692	&II	&$\mathrm{	5	h	12	m	}$&$\mathrm{	-69	^{\circ }	32	\arcmin	}$&	17 	&$\mathrm{	5	h	12	m	}$&$\mathrm{	-69	^{\circ }	32	\arcmin	}$&	29 	&	1.1 	&	0 	&	65	\\
247 &LMC N J0521-673	&I	&$\mathrm{	5	h	21.8	m	}$&$\mathrm{	-67	^{\circ }	42	\arcmin	}$&	17 	&$\mathrm{	5	h	22.1	m	}$&$\mathrm{	-67	^{\circ }	36	\arcmin	}$&	34 	&	0.9 	&	117 	&		\\
252 &LMC N J0528-672	&III	&$\mathrm{	5	h	28	m	}$&$\mathrm{	-67	^{\circ }	28	\arcmin	}$&	17 	&$\mathrm{	5	h	28	m	}$&$\mathrm{	-67	^{\circ }	28	\arcmin	}$&	29 	&	2.3 	&	0 	&		\\
253 &LMC N J0530-675	&I	&$\mathrm{	5	h	30.7	m	}$&$\mathrm{	-68	^{\circ }	0	\arcmin	}$&	17 	&$\mathrm{	5	h	31	m	}$&$\mathrm{	-67	^{\circ }	58	\arcmin	}$&	47 	&	2.1 	&	83 	&		\\
258 &LMC N J0535-661	&II	&$\mathrm{	5	h	35.3	m	}$&$\mathrm{	-66	^{\circ }	20	\arcmin	}$&	17 	&$\mathrm{	5	h	35.3	m	}$&$\mathrm{	-66	^{\circ }	18	\arcmin	}$&	24 	&	0.7 	&	28 	&		\\
266 &LMC N J0543-694	&II	&$\mathrm{	5	h	43.6	m	}$&$\mathrm{	-69	^{\circ }	48	\arcmin	}$&	17 	&$\mathrm{	5	h	43.5	m	}$&$\mathrm{	-69	^{\circ }	46	\arcmin	}$&	18 	&	3.6 	&	28 	&	207	\\
270 &LMC N J0547-695	&II	&$\mathrm{	5	h	48.1	m	}$&$\mathrm{	-70	^{\circ }	0	\arcmin	}$&	17 	&$\mathrm{	5	h	48.6	m	}$&$\mathrm{	-70	^{\circ }	6	\arcmin	}$&	82 	&	2.0 	&	143 	&	226,227	\\
271 &LMC N J0553-682	&I	&$\mathrm{	5	h	53.3	m	}$&$\mathrm{	-68	^{\circ }	24	\arcmin	}$&	17 	&$\mathrm{	5	h	53.3	m	}$&$\mathrm{	-68	^{\circ }	24	\arcmin	}$&	55 	&	1.9 	&	0 	&		\\

\enddata 

\tablenotetext{a}{Fukui et al.\ (2008)}
\tablenotetext{b}{Kawamura et al.\ (2009)}
\tablenotetext{c}{position of peak integrated intensity. Data is from Fukui et al.\ (2008). }
%\tablenotetext{d}{These radii were calculated from the equation of S=$\pi$R$^2$, and number of pixels detected with more than 80\% of the peak intensity. Asterisks show HI clouds that have a poorly defined extent. }
%\tablenotetext{d}{The radius was calculated from the equation of S=$\pi$R$^2$. S was calculated by multiplying (29 [pc])$^{2}$ (as one pixel) by the number of pixels detected with more than 80\% of peak integrated intensity. Asterisks show HI clouds that have a poorly defined extent. }
\tablenotetext{d}{HI cloud radius defined as $R = \sqrt{\frac{S}{\pi}}$. Here $S$ is the cloud area, calculated by summing the areas of all pixels detected above 80\% of the peak integrated intensity level. Asterisks show HI clouds whose extent is poorly defined.}
\tablenotetext{e}{HI column density of position at peak integrated intensity estimated by using the relation; $N$(HI) [cm$^{-2}$] = $1.82 \times 10^{18} \int T_b$(HI) $dv$ [K km s$^{-1}$]).}
\tablenotetext{f}{difference between CO peak position and HI peak position;\ (CO($\alpha, \delta$)-HI($\alpha, \delta$))}
\tablenotetext{g}{HI clouds including two or more GMCs; The numbers show GMCs that are located in the same HI cloud.}

\end{deluxetable}

%\end{document}

\clearpage

\begin{figure}
\epsscale{1.0}
\plotone{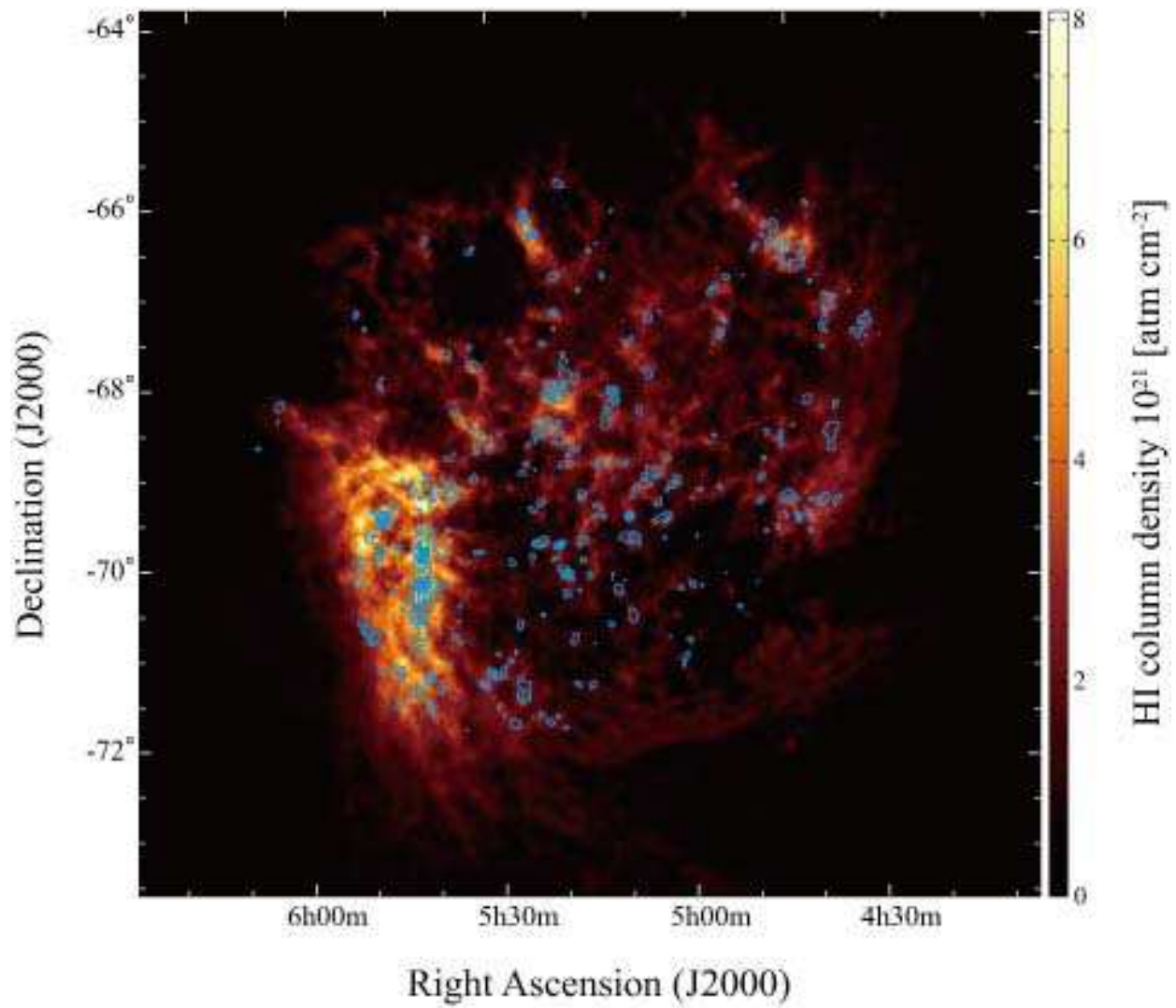}
\caption{ HI integrated intensity image (Kim et al. 2003) with
  contours of the CO integrated intensity (Fukui et al. 2008). The
  contour levels begin at 1.2 K km s$^{-1}$ and have 3.6 K km s$^{-1}$
  intervals.}
\label{fig:obsreg}
\end{figure}
\clearpage

\begin{figure}
\epsscale{0.7}
\plotone{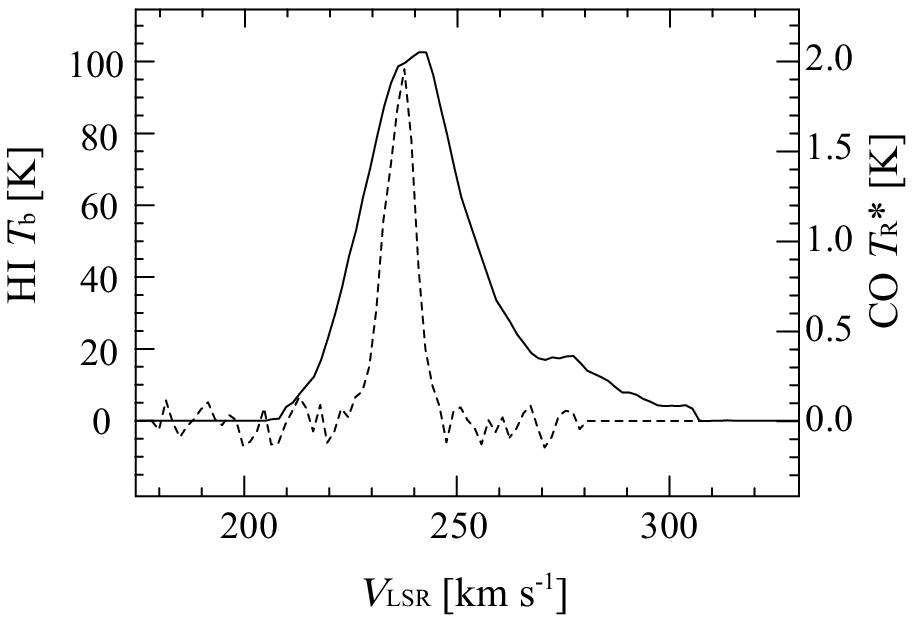}
\caption{An example of the HI and CO line profiles at $\alpha$(J2000) $=$ 5$^h$39$^m$38$^s$, and $\delta$(J2000) $= -69 \arcdeg 44\arcmin46\arcsec$.}
\label{fig:ii}
\end{figure}
\clearpage

\begin{figure}
\epsscale{0.6}
\plotone{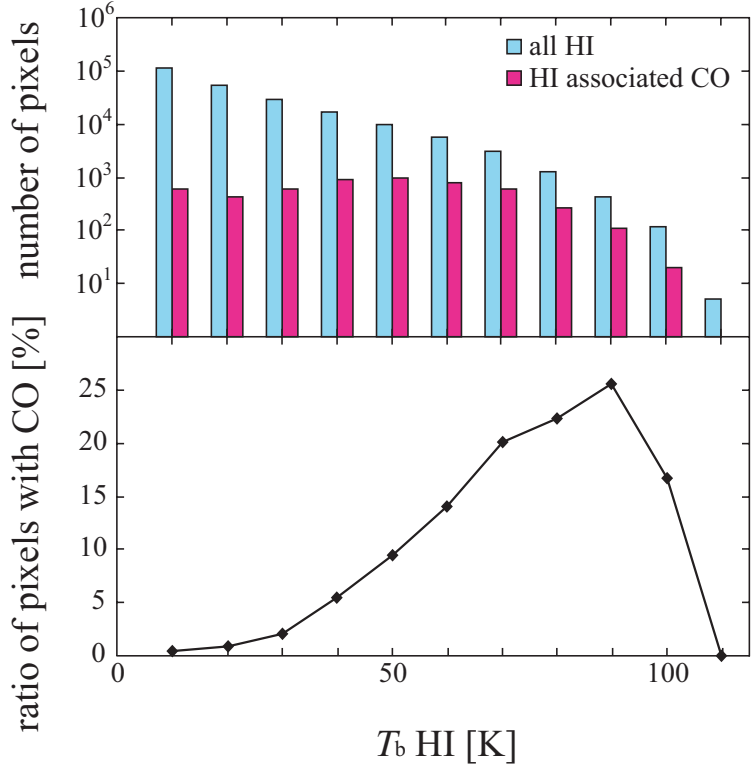}
\caption{The HI and CO datacubes that use for our analysis have
  somewhat different spatial and velocity resolutions; we have
  convolved both datasets to a spatial resolution of 40 pc $\times$ 40
  pc, and a velocity resolution of $\times$ 1.7 km s$^{-1}$ across the
  whole LMC. The pixels used in our analysis occupy the same volumes
  for both the HI and CO datasets. The $3 \sigma$ detection limits for
  the CO and HI brightness are 0.21 K and 7.2 K respectively. (a)
  Histogram of the HI brightness temperature averaged within each 3-D
  pixel is shown in blue; the HI pixels with significant CO emission
  (i.e. $> 3 \sigma$) are marked in red. The width of the histogram
  bins is 10 K. (b) Ratio of the number of pixels with significant HI
  emission to those with significant CO emission within each 10 K
  bin.}
\label{fig:iihist}
\end{figure}
\clearpage

\begin{figure}
\epsscale{1.0}
\plotone{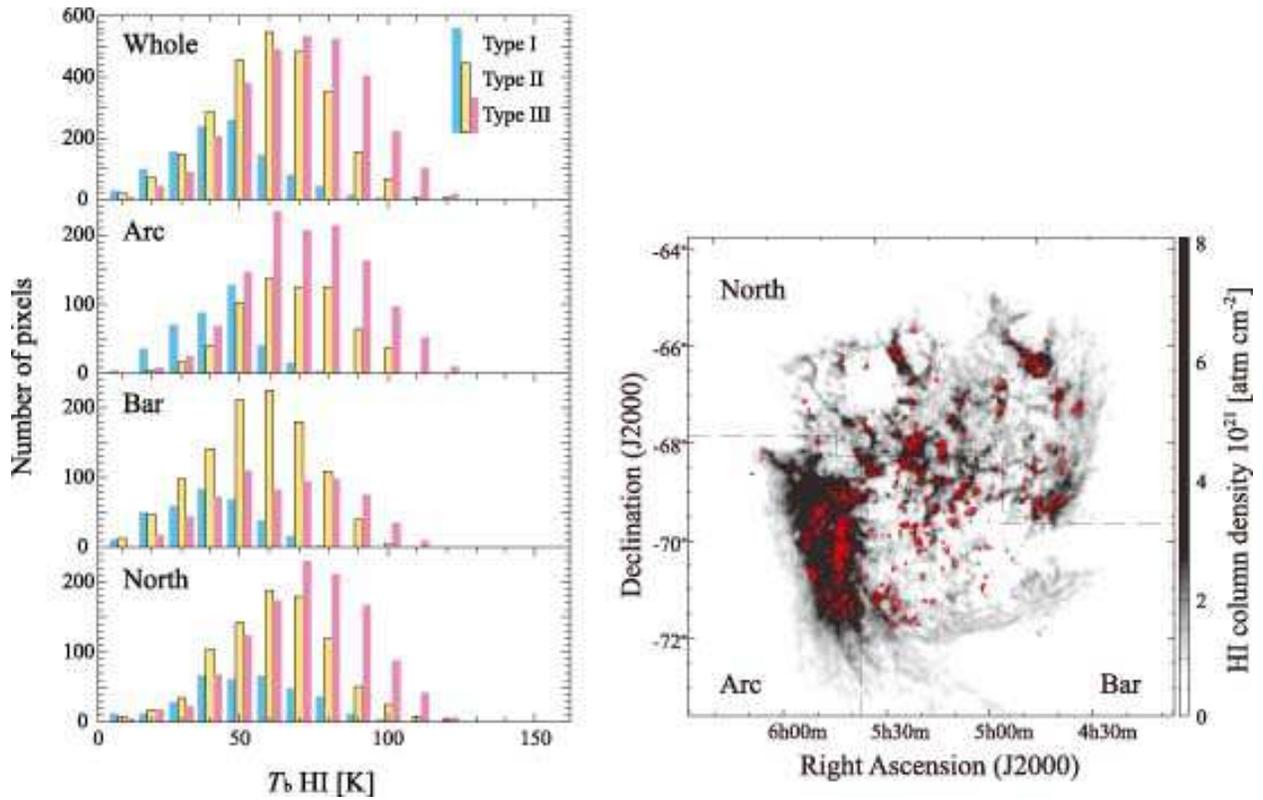}
\caption{Histograms of the pixel-averaged HI brightness temperature
  where significant CO emission is detected for Type I (blue), Type II
  (yellow) and Type III (red) GMCs. Histograms are shown for the whole
  LMC, and for three different regions -- Bar, North, and Arc -- which
  are indicated in the right panel.}
\label{fig:sddist}
\end{figure}
\clearpage

\begin{figure*}
\epsscale{1.0}
\plotone{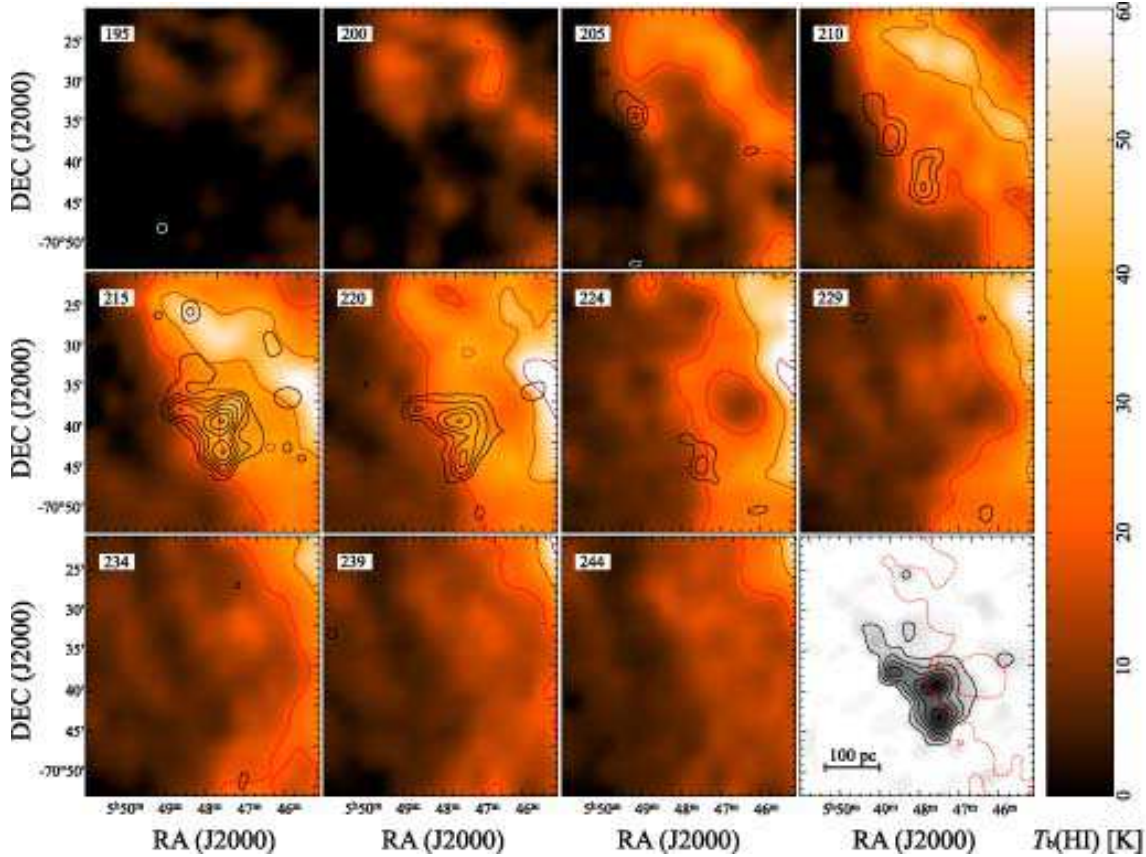}
\caption{Velocity channel maps of the HI and CO emission associated
  with Type I, Type II, and Type III, GMCs. The grayscale images with
  black contours are HI brightness temperature (Kim et al. 2003) and
  the white contours are CO intensity (Fukui et al. 2008). The central
  velocity of the included channels is shown in the upper left of each
  panel. The lower right panel of each figure shows the total CO
  integrated intensity. The red contours and the red cross in the
  lower right panel of (a) and (b) present the 80 \% level of the HI
  peak intensity and the position of the HI intensity peak,
  respectively. (a) GMC no. 225 in the NANTEN catalog. CO contour
  levels are from 0.105 K km s$^{-1}$ ($3 \sigma$) with 0.105 K km
  s$^{-1}$ intervals. HI contour levels are from 20 K with 20 K
  intervals. This is a case where the 80 \% level of the HI peak
  intensity is very extended relative to the size of the GMC. (b) GMC
  no. 134 and 135 between LMC4 and LMC5. CO integrated intensities are
  from 0.105 K km s$^{-1}$ with 0.105 K km s$^{-1}$ intervals. HI
  contour levels are from 20 K with 20 K intervals. This is a case
  where the 80 \% of the HI peak intensity provides a sensible
  definition of the HI envelope size. (c) GMC no. 192 and 202 with the
  HII region N159. CO integrated intensities are from 0.105 K km
  s$^{-1}$ with 0.21 K km s$^{-1}$ ($6 \sigma$) intervals. HI contour
  levels are from 20 K with 20 K intervals. This is a region where the
  HI emission is very complicated; it was not selected in the sample
  of 123 clouds that we use for our analysis.}
\label{fig:ch1}
\end{figure*}
\clearpage

\begin{figure*}
\epsscale{1.0}
{\plotone{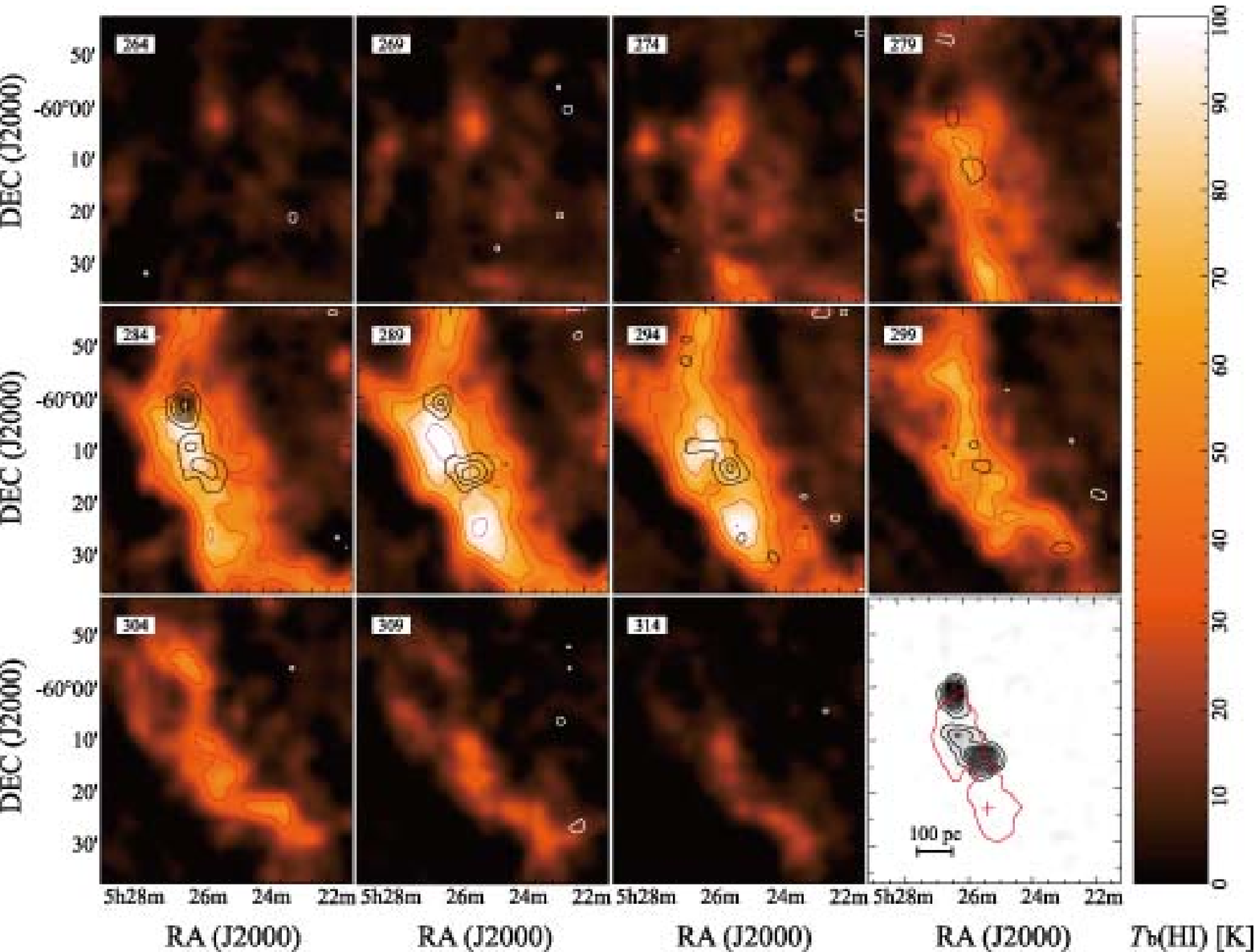}}\\[5mm]
\centerline{Fig. 5(b). --- {\it continued}}
\label{fig:ch2}
\end{figure*}
\clearpage

\begin{figure*}
\epsscale{1.0}
{\plotone{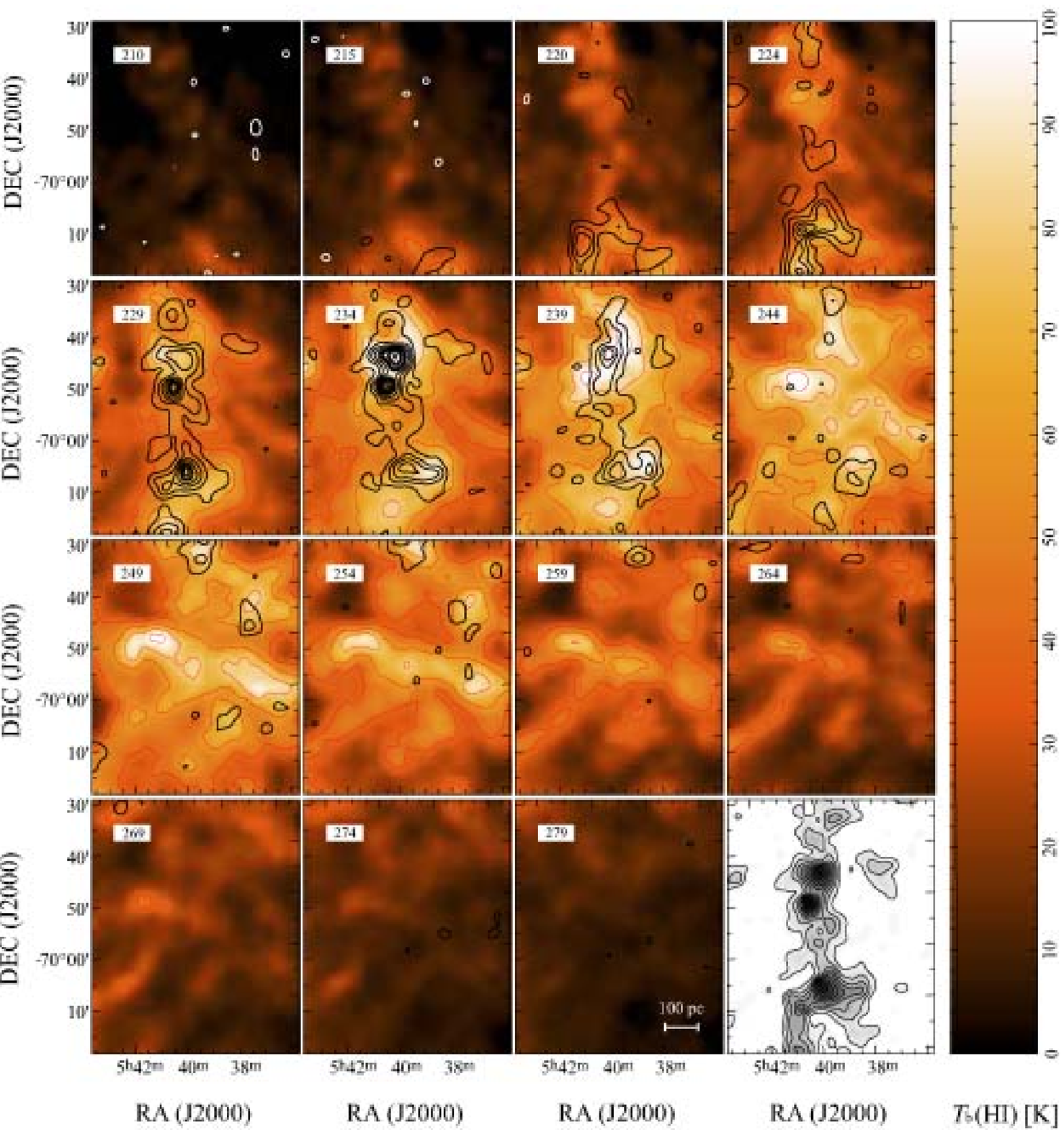}}\\[5mm]
\centerline{Fig. 5(c). --- {\it continued}}
\label{fig:ch3}
\end{figure*}
\clearpage

\begin{figure}
\epsscale{0.48}
\plotone{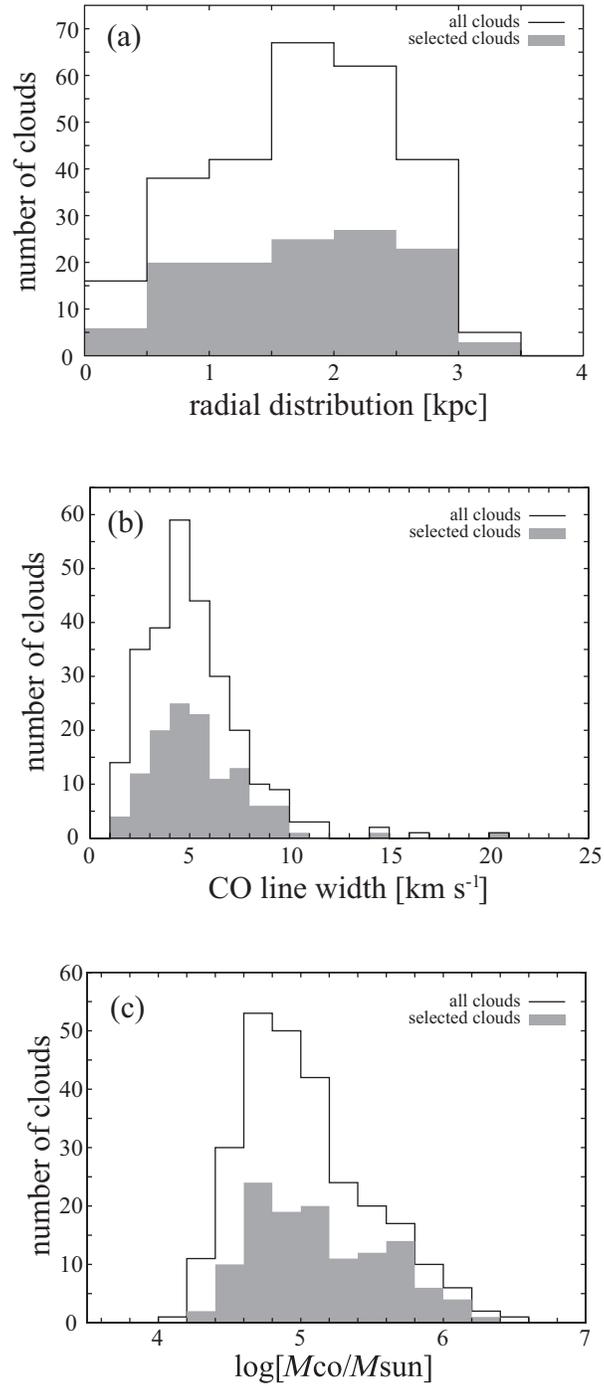}
\caption{ Histograms of three parameters, (a) the distance from the
  kinematic center of the galaxy (kpc), (b) the CO linwidth (kim
  s$^{-1}$) and (c) the CO cloud mass ($M_{\sun}$) for the 123 GMCs
  used in our analysis. The solid lines show the distribution of these
  properties for all the clouds in the Fukui et al. (2008) catalog;
  the gray shaded regions represent the distribution for the 123
  selected GMCs (see also Table 2). }
\label{fig:rms}
\end{figure}
\clearpage

\begin{figure}
\epsscale{0.5}
\plotone{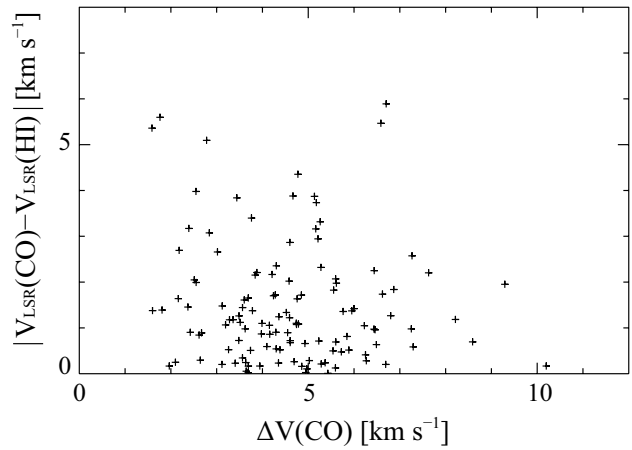}
\caption{The absolute difference between the central velocity of the
  HI and CO emission, $|V_{\rm LSR}{\rm (CO)}-V_{\rm LSR}{\rm (HI)}|$,
  versus the CO linewidth, $\Delta V$ (CO) for the 123 GMCs in our
  sample. The dotted line shows $|V_{\rm LSR}{\rm (CO)}-V_{\rm
    LSR}{\rm (HI)}| = \Delta V$(CO).  }
\label{fig:vlsrhist}
\end{figure}
\clearpage

\begin{figure}
\epsscale{0.5}
\plotone{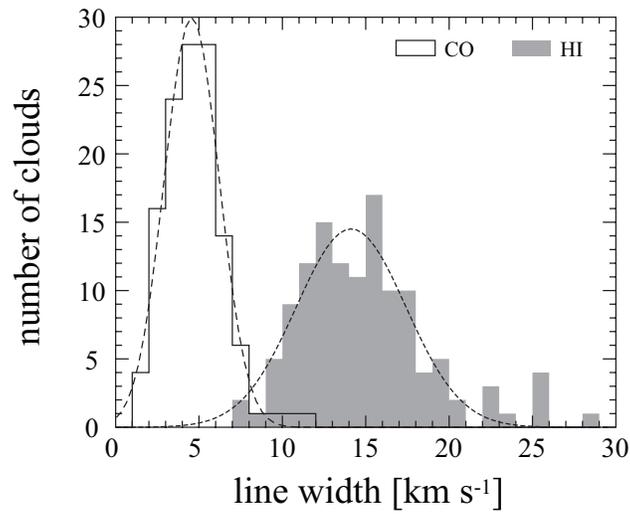}
\caption{ Histogram of the CO and HI linewidths toward the position of the CO peak
  position for the 123 selected GMCs. The dotted lines show
  Gaussian fits to the histograms. The mean values (standard derivations)
  for the CO and HI distributions are 4.6 (1.6) and 14.1 (3.3) km s$^{-1}$ respectively.
}
\label{fig:manmin}
\end{figure}
\clearpage

\begin{figure}
\epsscale{0.6}
\plotone{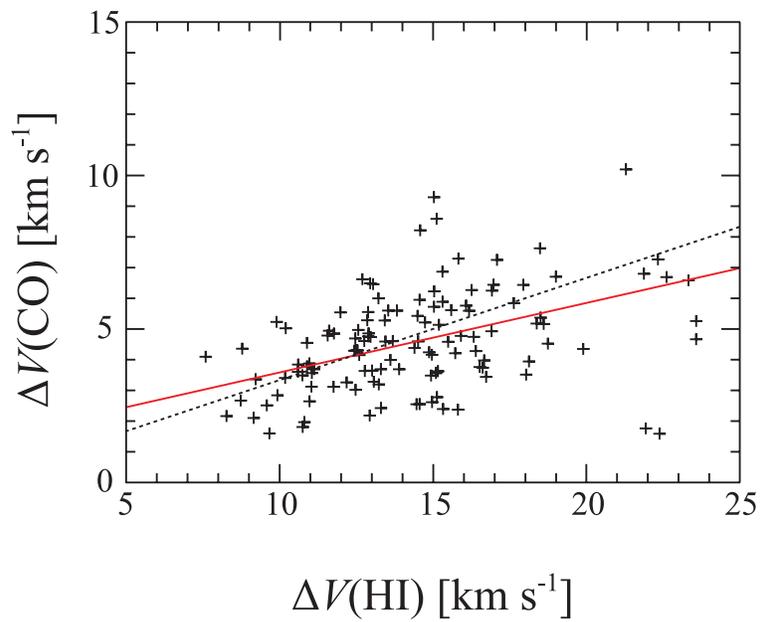}
\caption{ Plot of the CO versus HI linewidth for the 123 GMCs in our
  sample. The red line is the regression line $\Delta V{\rm (CO)} =
  {\rm (}1.32 \pm 0.04{\rm )}+{\rm(} 0.23 \pm 0.003{\rm)} \Delta V{\rm
    (HI)}$, and the dotted line shows $\Delta V{\rm (HI)} = 3 \times
  \Delta V{\rm (CO)}$. The Spearman rank correlation coefficient is 0.39. }
\label{fig:pa164}
\end{figure}
\clearpage

\begin{figure}
\epsscale{0.5}
\plotone{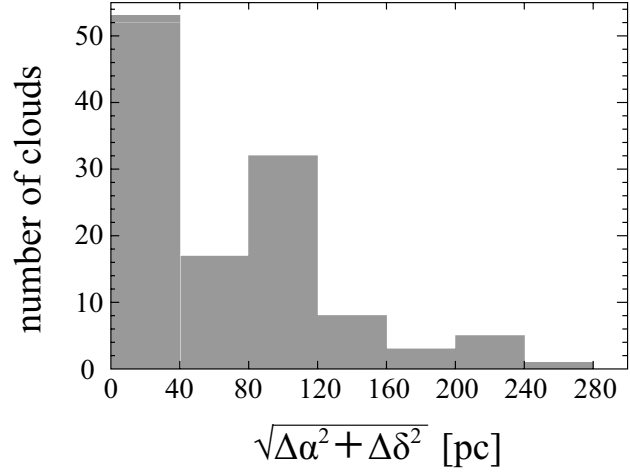}
\caption{ Histogram of the projected separation between the CO and HI
  peak positions, as listed Table 2. The separation between the CO and
  HI peak positions for four of the 123 clouds is greater than 300 pc;
  these clouds are not shown.  }
\label{fig:theta}
\end{figure}
\clearpage

\begin{figure}
\epsscale{0.6}
\plotone{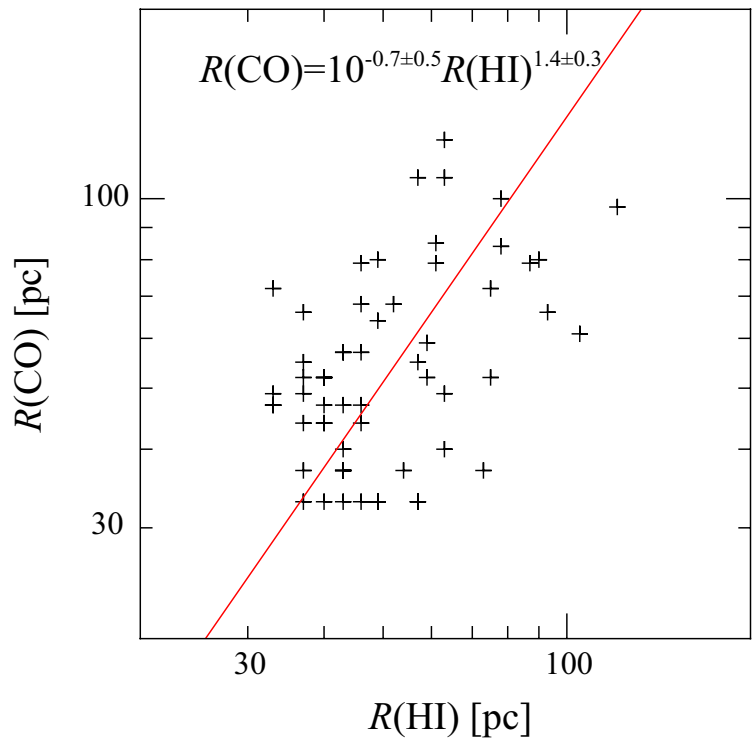}
\caption{ The radius of the GMC, $R$(CO), versus the radius of the HI
  envelope, $R$(HI) for the 62 clouds with radii greater than 30 pc,
  as listed in Table 2. The red line is the regression line $R$(CO)$=
  10^{-0.7 \pm 0.5} R$(HI)$^{1.4 \pm 0.3}$; the Spearman rank
  correlation coefficient is 0.45.}
\label{fig:rv}
\end{figure}

\begin{figure}
\epsscale{0.6}
\plotone{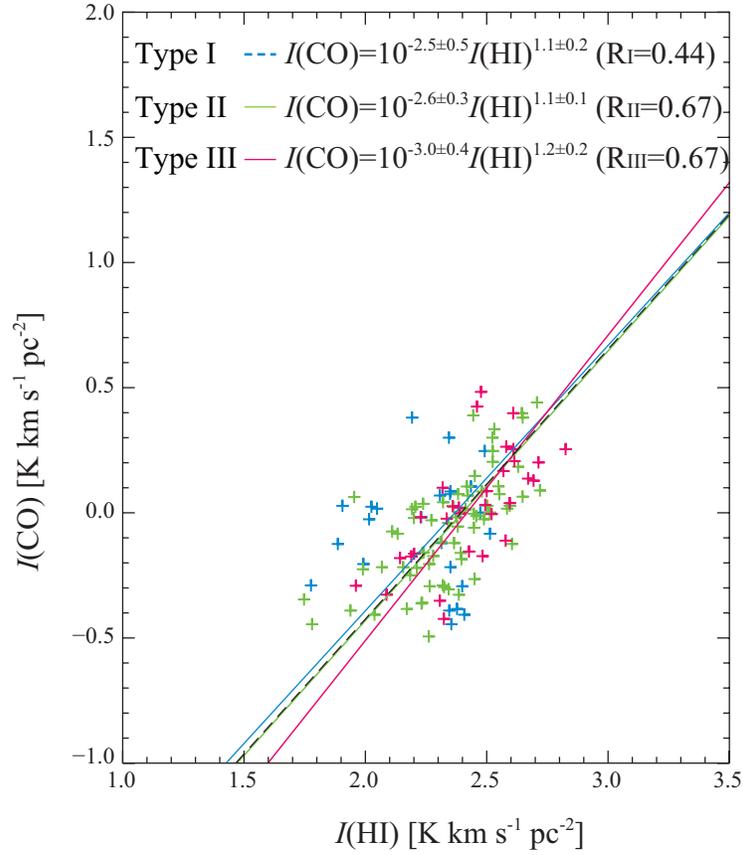}
\caption{ The CO versus HI integrated intensity for the 123 clouds in
  our sample, averaged over the projected area of each GMC or HI
  envelope. Blue, green and red plots symbols represent Type I, Type
  II and Type III GMCs respectively. The black dotted line shows the
  regression line for all 123 clouds; $I$(CO) $= 10^{-2.6 \pm
    0.2}$$I$(HI)$^{1.1 \pm 0.1}$. The regression lines and Spearman
  rank correlation coefficients (R$_{\rm I}$, R$_{\rm II}$, and
  R$_{\rm III}$) for the three GMC types are shown at the top of the
  plot. }
\label{fig:ico-ihi}
\end{figure}

\begin{figure}
\epsscale{0.7}
\plotone{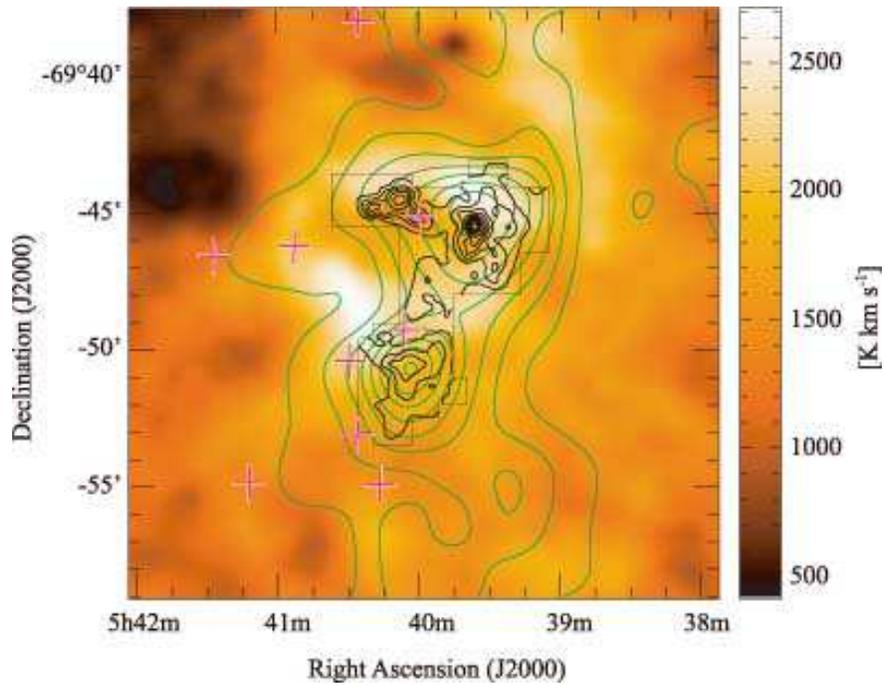}
\caption{ CO and HI emission in the N159 region.  The grayscale image
  represents the combined ATCA+Parkes HI integrated intensity data
  (Kim et al. 2003). The black contours are integrated CO ($J=$3--2)
  emission by ASTE (Minamidani et al. 2008) and green contours are
  integrated CO ($J=$1--0) emission by NANTEN (Fukui et al. 2008). For
  CO ($J=$3--2), the contour levels are from 5 K km s$^{-1}$ in 10 K
  km s$^{-1}$ intervals; for CO ($J=$1--0), the contours levels are
  from 1.2 K km s$^{-1}$ in steps of 2.4 K km s$^{-1}$. The thin black
  lines indicate the region observed in CO ($J=$3--2) emission by
  ASTE. The crosses indicate the positions of HII regions cataloged by
  Davies et al. (1976).  }
\label{fig:n159}
\end{figure}

\end{document}